\documentclass[journal,twoside]{IEEEtran}
\usepackage{ifpdf}
\usepackage{cite}
\usepackage{graphics} % for pdf, bitmapped graphics files
\usepackage{epsfig} % for postscript graphics files
\usepackage{mathptmx} % assumes new font selection scheme installed
\usepackage{times} % assumes new font selection scheme installed
\usepackage{amssymb}  % assumes amsmath package installed
\usepackage{multirow}
\usepackage{lipsum}
\usepackage{enumitem}

\usepackage{nomencl}

\usepackage [pdfstartview=FitH]{hyperref}

\usepackage{amsmath}
\usepackage{algorithmic}
\usepackage{array}
\ifCLASSOPTIONcompsoc
\usepackage[caption=false,font=normalsize,labelfont=sf,textfont=sf]{subfig}
\else
\usepackage[caption=false,font=footnotesize]{subfig}
\fi
\usepackage{fixltx2e}
\usepackage{dblfloatfix}
\ifCLASSOPTIONcaptionsoff
\usepackage[nomarkers]{endfloat}
\let\MYoriglatexcaption\caption
\renewcommand{\caption}[2][\relax]{\MYoriglatexcaption[#2]{#2}}
\fi
\let\MYorigsubfloat\subfloat
\renewcommand{\subfloat}[2][\relax]{\MYorigsubfloat[]{#2}}
\usepackage{url}

\begin{document}
\title{Robust $H_{\infty}$ Position Controller for Steering Systems}
\author{Tushar Chugh, Fredrik Bruzelius and Balázs Kulcsár% <-this % stops a space
	\thanks{This work is supported by the ITEAM project in the EU Horizon 2020 research and innovation program (grant no. 675999) and VINNOVA of the FFI, project SWOPPS (grant no. 2017–05504).}
	\thanks{Tushar Chugh is with Volvo Car Group, Gothenburg, Sweden (e-mail: tushar.chugh@volvocars.com). Fredrik Bruzelius and Balázs Kulcsár are with Chalmers University of Technology, Gothenburg, Sweden (e-mail: fredrik.bruzelius@chalmers.se and kulcsar@chalmers.se).}
}

\maketitle

\begin{abstract}
	This paper presents a robust position controller for electric power assisted steering and steer-by-wire force-feedback systems. A position controller is required in steering systems for haptic feedback control, advanced driver assistance systems and automated driving. However, the driver's \textit{physical} arm impedance causes an inertial uncertainty during coupling. Consequently, a typical position controller, i.e., based on single variable, becomes less robust and suffers tracking performance loss. Therefore, a robust position controller is investigated.
	
	The proposed solution is based on the multi-variable concept such that the sensed driver torque signal is also included in the position controller. The subsequent solution is obtained by solving the LMI$-H_{\infty}$ optimization problem. As a result, the desired loop gain shape is achieved, i.e., large gain at low frequencies for performance and small gain at high frequencies for robustness.
	
	Finally, frequency response comparison of different position controllers on real hardware is presented. Experiments and simulation results clearly illustrate the improvements in reference tracking and robustness with the proposed $H_\infty$ controller.
\end{abstract}


\section{Introduction}
\label{sec:introduction}
\IEEEPARstart{W}{ith} the advancement in driving automation technology, the transportation sector aims to realize semi-automated and automated vehicles. In the context of driving autonomy, there exists different levels of automation systems, depending on the undertaken \textit{dynamic driving task}, see~\cite{SAEJ30162018} for definitions. At operational level, to perform these tasks, both longitudinal and lateral vehicle motion should be controlled with a certain performance and safety guarantees~\cite{Magdici2017, Ni2016}. For lateral control, in the higher levels of driving automation especially in level $4$ and level $5$, the primary actuation must be done by the steering system to maintain the required trajectory. This task is executed by a steering position controller. Typically, it has a hierarchical structure: a higher level and a lower level control. The former computes an optimal reference trajectory, representing the vehicle motion control application. Whereas, the tracking is achieved with the latter by minimizing the error between the reference and actual position signals. For the higher level control, there has been an extensive research for more than a decade. For example using classical feedback-feedforward approach in~\cite{Kapania2015}, linear model predictive control (MPC) in~\cite{Ni2016}, piecewise linear MPC for the non-linear vehicle-tire interaction in~\cite{Cairano2010}, non-linear MPC method in~\cite{Falcone2007, Gao2014} to name a few. Most of them have reasonably assumed an ideal lower level control; however, we focus on that in this paper.

The higher level control requests a reference signal which can be either steering (or pinion/wheel) angle or rack position~\cite{Yang2015, Govender2016, Mardh2020}. In robotics, it is also referred to as admittance control, see e.g.~\cite{Aguirre-ollinger2012, Gupta2008, DeVlugt2003}. Regardless of the system dynamics, a position feedback controller minimizes the error in one of the interaction port's motion trajectory variables, i.e., in terms of angular position and/or velocity.

From the lateral vehicle control aspect, the maximum relevant frequency for periodic steering inputs is between $4$-$5$~Hz~\cite{VonGroll2006}, assuming no environmental disturbance and irrespective of the steering action performed by the driver or an advanced driver assistance function. This can be corroborated from the test scenarios considered in the higher level control design, see~\cite{Ni2016, Kapania2015, Falcone2007, Gao2014, Cairano2010}. Despite the fact, that most of the existing steering position controllers, for example in~\cite{Yang2015, Govender2016, Mardh2020}, would likely provide a good lateral vehicle control performance, the question is why do we seek another solution. The reason is coupling port's uncertainty that affects the nominal tracking.

\begin{figure}
	\centering
	\includegraphics[width=6.70cm]{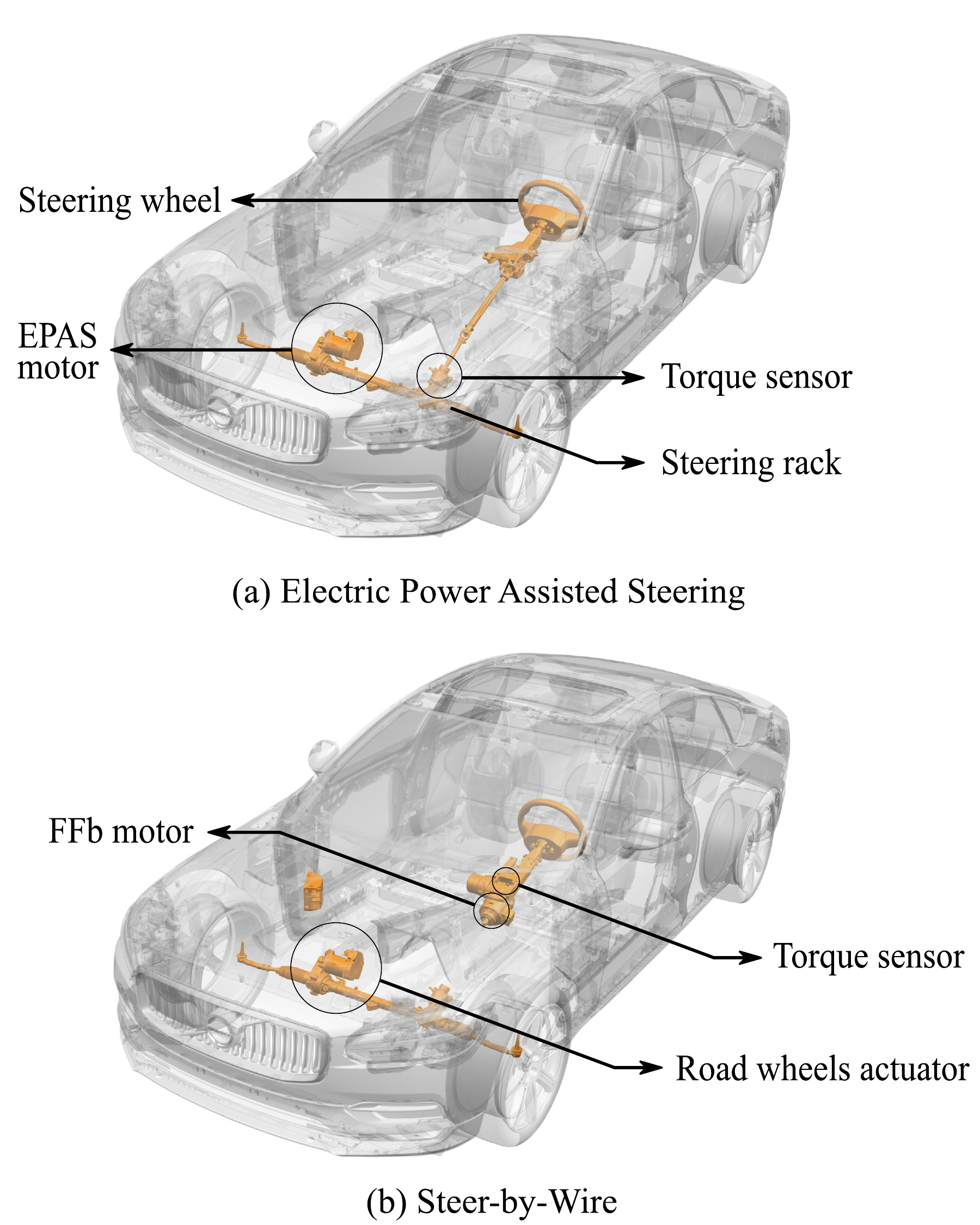}
	\caption{A typical illustration of: (a) an electric power assisted steering and (b) a steer-by-wire system.}
	\label{fig:SystemModel}
\end{figure}

In an electric power assisted steering (EPAS), there are two interaction ports: (a) the steering wheel responsible for the driver interaction; and (b) the steering rack responsible for the environment interaction, see Fig.~\ref{fig:SystemModel}(a). Although the driver is not required in level $5$ driving automation, but it is still a fail-safe fall back solution up to level $4$. As a result, the driver coupling could be a source of parametric uncertainty due to its own physical impedance, especially the coupled arm inertia. Whereas, an unknown tire-road interaction at the environment port (i.e. defined in terms of steering rack force) can be considered as a disturbance~\cite{Dannohl2012a}. For this, another possibility is to include a vehicle model at the environment port, thereby extending the state variables with: vehicle sideslip, yaw rate, and lateral tire slip angles, see e.g.~\cite{Huang2014, SadeghiKati2020}. In this approach, the vehicle and tire parameters, such as the peak tire-road friction coefficient, are assumed to be known; the model equations are also linearized with a limited tire model fidelity. Hence, this results in a real parametric uncertainty, such that the rack force becomes a function of vehicle speed and other parameters. Therefore,~\cite{Huang2014, SadeghiKati2020} have proposed a gain scheduled feedback controller based on some of the tire parameters and vehicle speed. To avoid these dependencies and for simplicity, we design a robust position controller regardless of the environmental interaction. %Irrespective of the approach, a closed-loop position controller should attenuate the rack force.

The steering feedback control is an important objective when the driver is in the loop. Hence, our goal is to use the same position controller for generating the required haptic feeling for consistency, unlike switching between the different controllers~\cite{Ott2010}. Consequently, an outer loop with a position reference generator is necessary for the closed-loop interaction, see~\cite{Hogan2004}. From the steering feedback perspective, the driver-environment transparency must be realized up to $15$~Hz, and so must the reference tracking performance~\cite{Chugh2021}.

For a typical steer-by-wire (SbW) system in Fig.~\ref{fig:SystemModel}(b), there are two actuators: (a) force-feedback (FFb) with the driver interaction port and (b) road wheel actuator with the environment interaction port. They are responsible for the steering feedback control and the vehicle lateral control objectives, respectively. A state-of-the-art road wheel actuator is based on position control such that the actual FFb angle defines the rack position reference. The road wheel actuator is disregarded in this paper because the environment port is mechanically disconnected from FFb and we assume no direct implication of this port's uncertainty on FFb~\cite{Wang2014, Sun2016, Sande2016}. However, the SbW-FFb system has the same tracking performance requirement as EPAS. 

In the literature, there are numerous examples of classical single variable position controllers, e.g.~\cite{Gualino2006, Zheng2019, Chugh2018a, Mardh2020}. They have limited robustness and a depreciating reference tracking performance during a port's uncertainty caused by the coupling impedance~\cite{Chugh2021}. Hence, there exists a research gap when it comes to a robust position controller for haptic applications. On the contrary, there are available torque/force controllers, for example~\cite{Gualino2007, Chugh2018a, Dannohl2012a}, where the feedback control principle is inherently robust against the coupling uncertainties due to its causality~\cite{Chugh2021}. Moreover, a further improvement was achieved in~\cite{Dannohl2012a} using the multi-variable approach by including the measured angular speed signal. 

The contribution of this paper is to include the sensed torque signal in the position control law for robustness. This is to mitigate the tracking performance loss due to the coupled port's parametric uncertainty. Thus, we investigate a multi-variable position controller using the $H_{\infty}$ framework for EPAS and SbW-FFb. The aim is to improve reference tracking, in case of an uncertainty, as compared to a classical solution. 

The system dynamics and the problem definition are described in Section~\ref{s:SystemDynamics}. The $H_\infty$ controller design and its results are presented in Section~\ref{s:RobustControl} and Section~\ref{s:Results}, respectively.

\section{System Dynamics and Problem Definition}
\label{s:SystemDynamics}
This section introduces the models for EPAS and SbW-FFb, as shown in Fig.~\ref{fig:SystemModel}. Furthermore, the problem is formulated.

\subsection{EPAS and SbW-FFb Modeling}
\label{s:Modeling}
\begin{figure*}
	\centering
	\includegraphics[width=17.75cm]{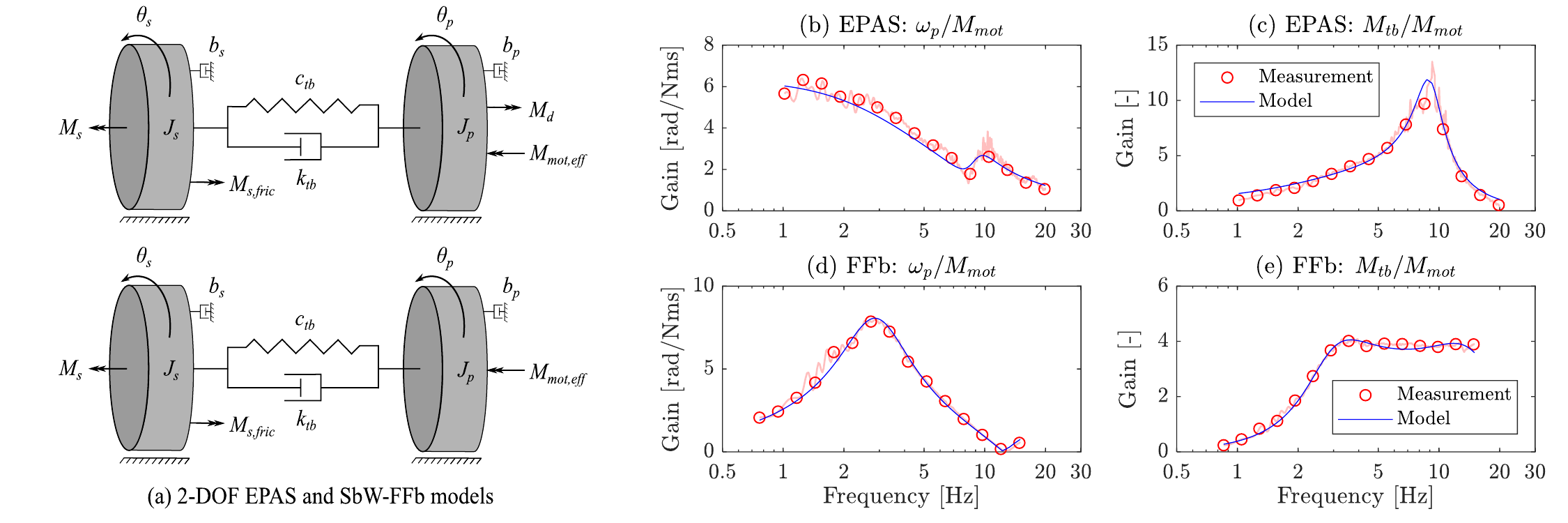}
	\caption{(a) Simplified $2$-DOF EPAS and SbW-FFb models, where $M_{mot,eff} = i_{mot} M_{mot}$. The system identification frequency response plots for EPAS: (b) $\omega_{p}(j\omega)/M_{mot}(j\omega)$, (c) $M_{tb}(j\omega)/M_{mot}(j\omega)$; and for FFb: (d) $\omega_{p}(j\omega)/M_{mot}(j\omega)$ and (e) $M_{tb}(j\omega)/M_{mot}(j\omega)$.}
	\label{fig:SystemIdentification}
\end{figure*}

Let us consider a model with $2$-DOF (degrees of freedom): steering angle, $\theta_{s}$, and pinion angle, $\theta_{p}$. The actuator is assumed to be rigidly connected to the steering column with an overall gear ratio, $i_{mot}$. In EPAS, the belt connection between the motor and steering rack can be translated to the pinion with the motor-to-rack and the rack-pinion gear ratios, such that $\theta_{p} = \theta_{mot}/i_{mot}$, where $\theta_{mot}$ is the motor angle. Whereas in SbW-FFb, the belt translation is directly from the motor to pinion. The belt stiffness is assumed sufficiently high for the controller synthesis. Hence, the equations of motion are:
\begin{eqnarray}
	\begin{split}
		J_{s} \dot{\omega}_{s}(t) &= -b_{s} \omega_{s}(t) - M_{tb}(t) + M_{s}(t) \\
		J_{p} \dot{\omega}_{p}(t) &= -b_{p} \omega_{p}(t) - M_{d}(t) + M_{tb}(t) + i_{mot} M_{mot}(t)
	\end{split}
	\label{eq:PlantDynamics}
\end{eqnarray}
where $M_{tb}(t) = c_{tb} (\theta_{s}(t) - \theta_{p}(t)) + k_{tb} (\omega_{s}(t) - \omega_{p}(t))$ is the sensed torsion bar torque, $M_{s}$ is the steering torque, $M_{d}$ is the disturbance torque (translated from the steering rack force), and $M_{mot}$ is the motor torque, see Fig.~\ref{fig:SystemIdentification}(a) for free-body diagrams. For FFb, $M_{d}$ is null due to mechanical disconnection. The Coulomb friction torques are excluded for a linear controller. During implementation, an observer was used to compensate for the motor friction torque, $M_{mot,fric}$.

During the driver-steering interaction, the human arms are coupled to the steering wheel. As a result, the system impedance is changed by the coupled arm inertia, $J_{arm}$, with the equation $J_{arm} \dot{\omega}_{s}(t) = M_{arm}(t) - M_{s}(t)$, where $M_{arm}$ is the arm torque. $J_{arm}$ could vary, depending on the driver and how firm the coupling is. Thus, it is a parametric uncertainty. Using these equations, a continuous-time state-space is formulated: 
\begin{eqnarray}
	\begin{split}
		\dot{x}(t) = A(t) x(t) + B_{1} d(t) + B_{2} u(t) \text{ and } y(t) = C x(t)
	\end{split}
	\label{eq:StateSpace}
\end{eqnarray}
such that $x(t) =[\theta_{s} \ \omega_{s} \ \theta_{p} \ \omega_{p}]^{T}$, $d(t) = [M_{arm} \ M_{d}]^{T}$, $u(t) = M_{mot}$, and $y(t) = [\omega_{s} \ \theta_{p} \ \omega_{p} \ M_{tb}]^{T}$. The Laplace representation is $Y_{i}(s) = [G_{i1} \ G_{i2} \ G_{i3}] [D(s) \ U(s)]^{T}$, where $i$ is the output channel, $G_{ij}$ is input $j$ to output $i$ transfer function, and $s$ is the Laplace operator. Also, $\omega_{i}(s) = s \theta_{i}(s)$ holds. %For the controller design, we will follow the output feedback approach and not the state feedback; because in reality there is a belt compliance and we measure $(\theta_{mot}, \omega_{mot})$, rather than $(\theta_{p}, \omega_{p})$.

\subsection{System Identification}
\label{s:SysId}
Parametric system identification procedure was performed to obtain the parameters. The systems were excited using the respective control actuator for realizing high frequencies.

In EPAS, the test was conducted on a prototype vehicle. The actuator was excited with a sine sweep $M_{mot}$ signal from $0.01$-$20$~Hz frequency and a peak amplitude of $0.4$ Nm. There was no steering input, i.e. $M_{s} = 0$ Nm or a free steering wheel. The rack force was measured using external strain gauges for validation. The vehicle speed was kept fairly constant at $10$ km/h for a consistent rack load. From the measured rack force signal, the equivalent disturbance torque was subtracted to form a single input channel. The measured signals were sampled at $1$ kHz. The measured and the fitted single input single output (SISO) frequency response functions (FRF) are shown in Fig.~\ref{fig:SystemIdentification}(b) and (c) for $G_{33}$ and $G_{43}$ respectively. The parameters in Table~\ref{t:ParametersDefinition} were estimated by minimizing the sum of normalized root mean square error (RMSE) in $G_{23}$, $G_{33}$ and $G_{43}$ between the measured and the model FRF using Equation~\eqref{eq:StateSpace}. For Fig.~\ref{fig:SystemIdentification}(b) and (c), the respective RMSE values are $0.456$ rad/s and $0.541$ Nm.

The FFb parameter identification was performed using the indirect closed-loop approach. The actuator was excited in a position control mode to obtain a high coherence at low frequencies. The reference pinion angle, $\theta_{p,ref}$, was excited with a sine sweep signal, $1^\circ$ peak amplitude from $0.01$-$15$~Hz and without any steering input. Again, a sampling rate of $1$ kHz was chosen. The parameters were identified by subtracting the contribution of the \textit{known} controller; and following the previous procedure. The results can be seen in Fig.~\ref{fig:SystemIdentification}(d)-(e) for the SISO measured and the model FRF in $G_{33}$ and $G_{43}$ respectively. The corresponding RMSE values are $0.123$ rad/s and $0.503$ Nm; see Table~\ref{t:ParametersDefinition} for the parameters.
\begin{table}
	\centering
	\caption{Definition \& Values of Model Parameters}
	\setlength{\tabcolsep}{3pt}
	\begin{tabular}{|p{25pt}|p{70pt}|p{45pt}|p{40pt}|p{35pt}|}
		\hline
		Symbol    & Description           & Value (EPAS) & Value (FFb)  & Units   \\
		\hline
		$b_{s}$   & Steering damping      & $0.1414$     & $0.0195$ 	& Nm-s/rad \\
		$J_{s}$   & Steering inertia      & $0.0337$ 	 & $0.0286$ 	& kgm$^2$ \\
		$c_{tb}$  & Torsion bar stiffness & $143.24$ 	 & $143.24$ 	& Nm/rad  \\
		$k_{tb}$  & Torsion bar damping   & $0.2292$ 	 & $0.1150$ 	& Nm-s/rad \\
		$b_{p}$   & Pinion damping        & $0.2964$ 	 & $0.0085$ 	& Nm-s/rad \\
		$J_{p}$   & Pinion inertia        & $0.1658$ 	 & $0.0017$ 	& kgm$^2$ \\
		$i_{mot}$ & Motor-pinion ratio    & $25^{*}$     & $3^{*}$      & $-$	  \\
		\hline
		\multicolumn{5}{p{240pt}}{$^*$Rounded-off due to proprietary reasons.}
	\end{tabular}
	\label{t:ParametersDefinition}
\end{table}

\subsection{Problem Formulation}
\label{s:ProbDefinition}
For defining the problem, consider a typical position control law in Equation~\eqref{eq:PosCtrlSingleVar}, where $e_{\theta}(t) = \theta_{p,ref}(t) - \theta_{p}(t)$. Previous work, in~\cite{Chugh2021}, shows Equation~\eqref{eq:PosCtrlSingleVar} sufficiently ensures hardware impedance compensation and reference tracking. Using Equations \eqref{eq:PosCtrlSingleVar} and \eqref{eq:PlantDynamics} to form the closed-loop interconnection.
\begin{eqnarray}
	\begin{split}
		M_{mot}(t) = \beta_{3} \ddot{e}_{\theta}(t) + \beta_{2} \dot{e}_{\theta}(t) + \beta_{1} e_{\theta}(t) + \beta_{0} \int_{0}^{t}e_\theta(t) d\tau
	\end{split}
	\label{eq:PosCtrlSingleVar}
\end{eqnarray}

The controller gains ($\beta_{0},\beta_{1},\beta_{2},\beta_{3}$) should be high for the performance reasons; except for the practical aspects such as time delay, quantization error, noise on the error states, etc. Therefore, we have used an analytical stability condition, from~\cite{Chugh2021}, in Equation~\eqref{eq:PosCtrlStab1}. It ensures no loop gain encirclement of $L(s) = K_{\theta}(s) G_{23}$ around ($-1,0$) in the Nyquist plane, where $K_{\theta}(s) = \beta_{3} s^2 + \beta_{2} s + \beta_{1} + \beta_{0}/s$ is the feedback control transfer function. Although $K_{\theta}(s)$ is acausal/improper, but each error state is assumed to be known for theoretical analysis.
\begin{eqnarray}
	\begin{split}
		\beta_{0} < \bigg(\frac{b_{s} + b_{p} + \beta_{2} i_{mot}}{J_{s} + J_{arm} + J_{p} + \beta_{3} i_{mot}}\bigg) \beta_{1}
	\end{split}
	\label{eq:PosCtrlStab1}
\end{eqnarray}
The tracking performance is evaluated by $\theta_{p}/\theta_{p,ref}$ FRF. Assuming a low-pass filter behavior, this FRF defines the controller bandwidth with cut-off frequency $\omega_{c}$. The uncertainty problem is presented by considering the percentage decrease in $\omega_{c}$ with respect to $\omega_{c_{0}}$, i.e. the nominal case with no uncertainty. The controller gains in Table~\ref{t:ControlGains} ensure $60^\circ$ and $40^\circ$ nominal phase margin for EPAS and FFb respectively.
\begin{table}
	\centering
	\caption{Real-Time Controller Gains}
	\setlength{\tabcolsep}{3pt}
	\begin{tabular}{|p{30pt}|p{50pt}|p{50pt}|p{50pt}|}
		\hline
		Symbol      	 & Value (EPAS) & Value (FFb) & Units        \\
		\hline
		$\beta_{0}$ 	 & $8$          & $15$ 	      & Nm/rad/s	 \\
		$\beta_{1}$ 	 & $5$          & $5$ 	      & Nm/rad	     \\
		$\beta_{2}$  	 & $0.48$       & $0.325$     & Nm-s/rad	     \\
		$\beta_{3}$  	 & $0.0065$     & $0.00035$   & Nms$^2$/rad  \\
		$\alpha_{0}$ 	 & $-6$         & $-15$	      & $1$/s		 \\
		$\alpha_{1}$ 	 & $-0.35$      & $-0.145$    & $-$      	 \\
		$\alpha_{1}'$ & $-0.0175$    & $0.0725$    & $-$      	 \\
		\hline
	\end{tabular}
	\label{t:ControlGains}
\end{table}

\begin{figure}
	\centering
	\includegraphics[width=9cm]{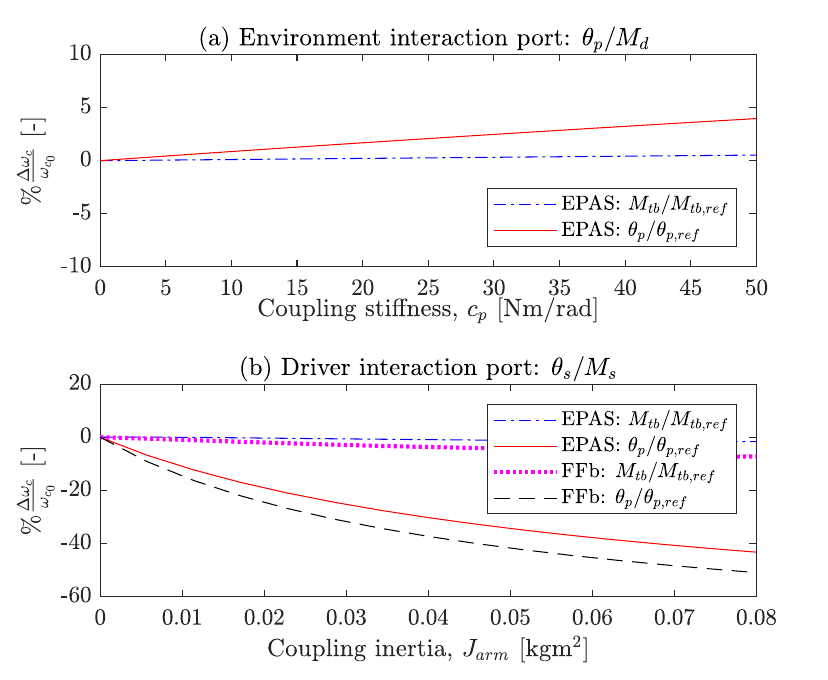}
	\caption{A relative \% change in bandwidth due to: (a) coupling stiffness $c_{p}$ at $\theta_{p}/M_{d}$ port and (b) coupling inertia $J_{arm}$ at $\theta_{s}/M_{s}$ port.}
	\label{fig:ProbDefinition}
\end{figure}

For the environment interaction port's uncertainty in EPAS, $M_{d}(t) = c_{p} \theta_{p}(t)$ is considered, where $c_{p}$ is the coupling or the disturbance stiffness. As shown in Fig.~\ref{fig:ProbDefinition}(a), $\theta_{p}/\theta_{p,ref}$ tracking is not affected much within a reasonable $c_{p}$ variation; because $\% \Delta \omega_{c}/\omega_{c_{0}}$ changes marginally. Although there exists an upper bound on $c_{p}$ after which the performance would drop, but those high $c_{p}$ values are kept out of context. Therefore, the vehicle and the tire models are avoided here for a less complicated solution.

The driver interaction port's uncertainty originates from the coupled arm inertia. The relative controller bandwidth reduction is depicted in Fig.~\ref{fig:ProbDefinition}(b) as a function of $J_{arm}$, using Equation~\eqref{eq:PosCtrlSingleVar} control law. The performance loss with high $J_{arm}$ would result in a poor tracking at high frequencies. 

For comparison, the results from~\cite{Chugh2021} using a Proportional-Integrator torque controller are also provided. The control signal in Laplace domain is $M_{mot}(s) = (\alpha_{1} + \alpha_{0}/s) e_{M}(s)$, where $e_{M}(s) = M_{tb,ref}(s) - M_{tb}(s)$; such that the tracking is defined by $M_{tb}/M_{tb,ref}$ FRF. As it can be seen in Fig.~\ref{fig:ProbDefinition}(b), the torque control method does not suffer any performance loss by $J_{arm}$ coupling. This is another motivation for investigating a comparable robust position controller.

To increase the position control performance, we require either a powerful actuator i.e. a high $M_{mot}$ limit and/or fast actuation through high $\beta_{2}$, $\beta_{3}$ values for high load inertia. Both the solutions are limited due to the design factors or the signal latency and/or error state noise levels respectively.

This paper investigates a solution on how to mitigate the tracking performance drop using the robust control design method by reformulating the position control law.

\section{Robust Position Control Design}
\label{s:RobustControl}
In the robust feedback control approach, we include the measured torque $M_{tb}$ along with the conventional pinion angle error states $e_{\theta}(t)$, $\dot{e}_{\theta}(t)$ and/or $\ddot{e}_{\theta}(t)$. The remainder of this section will introduce two steps with an increasing complexity: (a) a position control with a static torque feedback gain and (b) an $H_{\infty}$ controller solvable by the linear matrix inequality (LMI) methods.

\subsection{Position Control with Static Torque Feedback}
\label{s:MultVarPosCtrl}
A straightforward proposition is to include $M_{tb}(t)$ with a linear static gain in Equation~\eqref{eq:PosCtrlSingleVar}. Thus, a multi-variable feedback control law is formed, such that $M_{mot}(t) =$
\begin{eqnarray}
	\begin{split}
		\beta_{3} \ddot{e}_{\theta}(t) + \beta_{2} \dot{e}_{\theta}(t) + \beta_{1} e_{\theta}(t) + \beta_{0} \int_{0}^{t}e_\theta(t) d\tau + \alpha_{1}' M_{tb}(t).
	\end{split}
	\label{eq:PosCtrlMultiVar}
\end{eqnarray}
Again, the closed-loop interconnection is formed using the pinion's equation of motion and Equation~\eqref{eq:PosCtrlMultiVar}. The resulting loop gain becomes $L(s) = K_{\theta}(s) G_{23}/(1 - K_{M}(s) G_{43})$, where $K_{\theta}(s) = \beta_{3} s^2 + \beta_{2} s + \beta_{1} + \beta_{0}/s$ and $K_{M}(s) = \alpha_{1}'$. The stability conditions are provided below derived using the Simplified Nyquist criterion with the same assumptions as before. The first condition has evolved from Equation~\eqref{eq:PosCtrlStab1} due to $\alpha_{1}'$. With increasing $\alpha_{1}' (> 0)$, maximum $\beta_{0}$ bound will decrease and vice versa for $\alpha_{1}' < 0$. On the other hand, the second inequality constraint gives a minimum bound on $\alpha_{1}'$.
\begin{eqnarray}
	\begin{split}
		\beta_{0}      &< \bigg(\frac{b_{s} + \frac{b_{p} + \beta_{2} i_{mot}}{1 + \alpha_{1}' i_{mot}}}{J_{s} + J_{arm} + \frac{J_{p} + \beta_{3} i_{mot}}{1 + \alpha_{1}' i_{mot}}}\bigg) \beta_{1} \\
		\alpha_{1}' &> - \frac{1}{i_{mot}} \bigg[1 + \frac{b_{s}}{b_{p} + \beta_{2} i_{mot}} \bigg( \frac{J_{p} + \beta_{3} i_{mot}}{J_{s} + J_{arm}} \bigg)^{2} \bigg]
	\end{split}
	\label{eq:PosCtrlStab2}
\end{eqnarray}
Here, the new loop transfer function $L_{1}(s)$ is modified by the term $\Delta(s) = (1 - K_{M}(s) G_{43})^{-1}$ in comparison to the classical position control case. For simplicity, we rewrite the following: $L_{1}(s) = L_{0}(s) \Delta (s)$, such that $L_{0}(s) = K_{\theta}(s) G_{23}$ is the nominal loop gain; see Appendix~\ref{app:Definition} for the definitions.

In the upcoming steps, the importance of torque feedback from the performance and the robustness perspective is explained. For simplicity, the analytical expression for $\Delta(s)$ can be resolved by assuming the \textit{ideal} driver model condition. It implies a fixed steering wheel and decoupling of the steering wheel equation of motion from the pinion dynamics. Thus, $(\theta_{s}, \omega_{s}) = (0,0)$ is considered. Hence, $\Delta(s)$ is given in Equation~\eqref{eq:LoopGainPosCtrl} with $k_{tb} = 0$ Nm/rad/s. For understanding, the uncoupled $\Delta(s)$, i.e. for a free steering wheel motion, is given in Equation~\eqref{eq:UncoupledLoopGain} in Appendix~\ref{app:LoopGain}.

From the literature and $\Delta(s)$ below, it is intuitive that the pinion's equation of motion is altered by $(1 + \alpha_{1}' i_{mot})^{-1}$. 
\begin{eqnarray}
	\begin{split}
		\Delta(s) &= \frac{J_{p} s^2 + b_{p} s + c_{tb}}{J_{p} s^2 + b_{p} s + c_{tb} (1 + \alpha_{1}' i_{mot})} \\
	\end{split}
	\label{eq:LoopGainPosCtrl}
\end{eqnarray}
$G_{c}(s) = L(s)/(1+L(s))$ defines the tracking performance. The nominal loop gain $L_{0}(s)$ could be preserved for $L_{1}(s)$ by modifying $K_{\theta}(s) \to K_{\theta}(s)/\Delta(s)$ for the same tracking performance and the robustness. A better performance requires a large loop gain, but certainly with a reduced robustness.

\textit{Proposition 3.1:} With $\Delta(j \omega)$ in Equation~\eqref{eq:LoopGainPosCtrl} and $\alpha_{1}' < 0$, such that $\alpha_{1}' \in (-1/i_{mot},0)$, the tracking performance could be improved without compromising the nominal phase margin.

The low and the high frequency response of $\Delta(j \omega)$ are derived below, using the initial and the final value theorems.
\begin{eqnarray}
	\begin{split}
		\lim\limits_{\omega \rightarrow 0} \Delta(j \omega) = \frac{1}{1 + \alpha_{1}' i_{mot}} = \Delta_{0} \text{ and } \lim\limits_{\omega \rightarrow \infty} \Delta(j \omega) = 1
	\end{split}
	\label{eq:Lemma1a}
\end{eqnarray}
In case of a positive torque feedback, i.e. $\alpha_{1}' > 0$, $\Delta(j \omega)$ has a phase-lead character. Because $|L_{1}| = |L_{0}| \Delta_{0} < |L_{0}|$ in the low frequency region (as $\omega \rightarrow 0$); but retracts to its nominal shape for $\omega \rightarrow \infty$, such that $|L_{1}| = |L_{0}|$ at high frequencies. Therefore, the tracking performance drops with an increasing $\alpha_{1}'$. A sufficient condition $\omega_{\phi_{m}} \gg \sqrt{c_{tb} (1 + \alpha_{1}' i_{mot}) /J_{p}}$ guarantees $L_{1}(s)$ retains the nominal phase margin, where  $\omega_{\phi_{m}}$ is $L_{0}(s)$ gain crossover frequency. For a very stiff system, i.e. $c_{tb} \rightarrow \infty$, the controller gains $\beta_{3}, \beta_{2}, \beta_{1}$ and $\beta_{0}$ should be increased by $1/\Delta_{0}$ to preserve the nominal loop gain shape for the same performance and the robustness.

However with $\alpha_{1}' < 0$, $\Delta(j \omega)$ exhibits a phase-lag compensatory behavior. This implies, a larger loop gain at low frequencies for a higher bandwidth, i.e. $|L_{1}| = |L_{0}| \Delta_{0} > |L_{0}|$ using Equation~\eqref{eq:Lemma1a}. And $|L_{1}| = |L_{0}|$ at high frequencies, thus ensuring the nominal phase margin, given $\omega_{\phi_{m}} \gg \sqrt{c_{tb} / J_{p}}$. Hence, we could improve the tracking with a minimal influence on the phase margin using a negative torque feedback. 

For a free steering wheel, there exists a maximum bound on $\alpha_{1}'$ beyond which the phase margin would eventually drop due to the coupling effect from the steering equation of motion in $\Delta(s)$. But the above proposition still holds within that specific $\alpha_{1}'$ region even without a driver coupling.

With a positive torque feedback, the performance is decreased because of a smaller loop gain at low frequencies, i.e. $|L_{1}(j \omega)| < |L_{0}(j \omega)|$. Consequently, $K_{\theta}$ should be increased to preserve the existing loop gain shape. Whereas in a negative torque feedback for an infinitely stiff system, the existing $K_{\theta}$ is inherently amplified by $\Delta_{0} > 1$, such that the control actuation becomes $|K_{\theta}(j \omega)| \Delta_{0} > |K_{\theta}(j \omega)|$. For experiments, it means $\beta_{3}, \beta_{2}, \beta_{1} \text{ and } \beta_{0}$ are kept unchanged from the previous section, in addition to $\alpha_{1}' < 0$.

We are ready to investigate if a static torque feedback gain with $\alpha_{1}' < 0$ is ``optimal'' enough or not, in terms of tracking objectives and robustness.

\textit{Proposition 3.2:} The tracking improves with $\alpha_{1}' < 0$ or $\Delta_{0} > 1$, given \textit{Proposition 3.1} is fulfilled, but it reduces the closed-loop damping ratio, thus causing an overshoot.

Lets assume the nominal tracking transfer function, with Equation~\eqref{eq:PosCtrlSingleVar} as the control law, is given by a $1^{\text{st}}$-order function: $G_{c_{0}}(s) = L_{0}(s)/(1 + L_{0}(s)) = (1 + T_{0} s)^{-1}$, where $\omega_{0} = 1/T_{0}$ defines the cut-off frequency. Consequently, the nominal loop gain becomes $L_{0}(s) = (T_{0} s)^{-1}$.

With torque feedback $\alpha_{1}' < 0$, $\Delta(s)$ in Equation~\eqref{eq:LoopGainPosCtrl} and $L_{1}(s) = L_{0}(s) \Delta(s) = \Delta(s) (T_{0} s)^{-1}$, the updated tracking transfer function becomes: $G_{c_{1}}(s) = L_{1}(s)/(1 + L_{1}(s)) =$
\begin{eqnarray}
	\begin{split}
		\frac{J_{p} s^2 + b_{p} s + c_{tb}}{J_{p} T_{0} s^3 + (J_{p} + b_{p} T_{0}) s^2 + (b_{p} + c_{tb} T_{0} (1 + \alpha_{1}' i_{mot})) s + c_{tb}}.
	\end{split}
	\nonumber
\end{eqnarray}
$G_{c_{1}}(j \omega)$ can be solved to verify if there exists any valid interval for $\omega$, such that $|G_{c_{1}}(j \omega)| > 1$ holds. As a reasonable simplification, we assume $J_{p} \gg b_{p} T_{0}$. Thus, upon solving, Equation~\eqref{eq:Lemma2a} is derived.
\begin{eqnarray}
	\begin{split}
		\underbrace{\sqrt{\frac{c_{tb} (1 + \alpha_{1}' i_{mot})}{J_{p}}}}_{\underline{\omega}_{1}} < \omega < \underbrace{\sqrt{\frac{c_{tb} (1 + \alpha_{1}' i_{mot}) + \frac{2 b_{p}}{T_{0}}}{J_{p}}}}_{\overline{\omega}_{1}}
	\end{split}
	\label{eq:Lemma2a}
\end{eqnarray}
$||G_{c_{1}}(j \omega^{\star})||_{\infty}$ defines the maximum value for $\omega^{\star} \in (\underline{\omega}_{1}, \overline{\omega}_{1})$. It is approximately given as following:
\begin{eqnarray}
	\begin{split}
		||G_{c_{1}}(j \omega^{\star})||_{\infty} \approx \sqrt{1 + \frac{b_{p}^2 \big(\underline{\omega}_{1}^2 + \frac{b_{p}}{J_{p} T_{0}} \big)}{\big(c_{tb} \alpha_{1}'i_{mot} + \frac{b_{p}}{T_{0}}\big)^2}} > 1.
	\end{split}
	\label{eq:Lemma2b}
\end{eqnarray}
Given $\alpha_{1}' < -b_{p}/(T_{0} c_{tb} i_{mot})$, although the damping ratio increases because $||G_{c_{1}}||_{\infty}$ decreases with decreasing $\alpha_{1}'$, but an overshoot still occurs. Hence, $|G_{c_{1}}(j \omega)| > 1, \forall \ \omega \in (\underline{\omega}_{1}, \overline{\omega}_{1})$. This proves the existence of an overshoot with respect to a given nominal $1^{\text{st}}$-order tracking response, i.e. defined by $T_{0}$. It is caused by the torsional eigenmode of the sensor; and this might not be desirable for an ideal reference tracking.

\textit{Proposition 3.3:} A requirement for an improvement in the nominal phase margin is to at least have a $2^{nd}$-order torque feedback, regardless of $\alpha_{1}'$.

Lets assume an acausal torque feedback transfer function is given by $K_{M}(s) = \alpha_{3}' s^2 + \alpha_{2}' s + \alpha_{1}'$, along with $K_{\theta}(s)$. Basically, it implies a manipulation through higher orders of the torque signal.

Recomputing $\Delta(s)$ with the above definition of $K_{M}(s)$ and then, using Equation~\eqref{eq:Lemma1a} for the high frequency response; given $\omega_{\phi_{m}} \gg \sqrt{c_{tb} / J_{p}}$, the following can be deduced.
\begin{eqnarray}
	\begin{split}
		\lim\limits_{\omega \rightarrow \infty} \Delta(j \omega) = \frac{J_{p}}{J_{p} + c_{tb} \alpha_{3}' i_{mot}} = \underline{\Delta}
	\end{split}
	\label{eq:Lemma3a}
\end{eqnarray}
A small loop gain requires $\underline{\Delta} < 1$. Hence for a higher phase margin at a given $\omega^{\star}$, $|L_{1}(j \omega^{\star})| < |L_{0}(j \omega^{\star})| \iff \alpha_{3}' > 0$, regardless of $K_{\theta}(j \omega)$ and $\alpha_{1}'$.

The question is, \textit{why a small loop gain at high frequencies is required other than robustness reasons}. The answer lies in the closed-loop formulation of plant, controller, and haptic feedback reference models. From the literature~\cite{Colgate1988, Chugh2021}, it is established that both uncoupled and coupled position control stability depends on the reference inertia, $J_{ref}$. 

\textit{Proposition 3.4:} To realize a high frequency steering feedback reference in the position control setting, the reference inertia must be small. However, it has a lower bound for ensuring the driver uncoupled stability.

To show this mathematically, consider a position reference function, i.e. $H_{ref}(s)$, assuming the reference stiffness and damping parameters to be null. Thus, the result is as follows.
\begin{eqnarray}
	\begin{split}
		H_{ref}(s) = \frac{\theta_{p,ref}(s)}{M_{tb}(s)} = \frac{1}{J_{ref}s^2}
	\end{split}
	\label{eq:ReferenceFcn}
\end{eqnarray}

Due to $H_{ref}(s)$ outer loop interconnection, there is another loop gain that defines the overall stability. Here, we illustrate how $J_{ref}$ depends on the inner loop gain. The overall loop gain, $L_{h}(s)$, is derived in Equation~\eqref{eq:OuterLoopGain1a} using Equation~\eqref{eq:StateSpace}, where $L_{i}(s)$ is the inner loop gain. This analysis is valid for an uncoupled driver port, but $J_{arm}$ is included as a plant parameter.
\begin{eqnarray}
	\begin{split}
		L_{h}(s) &= - \frac{M_{tb}(s)}{\theta_{p,ref}(s)} H_{ref}(s) = -G_{c} \frac{G_{43}}{G_{23}} H_{ref}(s) \\
				 &= \frac{L_{i}(s)}{1 + L_{i}(s)} \frac{((J_{s} + J_{arm})s^2 + b_{s}s) c_{tb}}{(J_{s} + J_{arm})s^2 + b_{s}s + c_{tb}} \frac{1}{J_{ref}s^2}
	\end{split}
	\label{eq:OuterLoopGain1a}
\end{eqnarray}
$L_{h}(s)$ can be further simplified by assuming $b_{s} = 0$ Nm-s/rad and $c_{tb} \to \infty$. Thus, the resulting expression is as follows.
\begin{eqnarray}
	\begin{split}
		L_{h}(s) &\approx \frac{L_{i}(s)}{1 + L_{i}(s)} \frac{J_{s} + J_{arm}}{J_{ref}} = G_{c_{i}}(s) \frac{J_{s} + J_{arm}}{J_{ref}}
	\end{split}
	\label{eq:OuterLoopGain1b}
\end{eqnarray}
Applying the small gain theorem to $L_{h}(s)$ with an implication: $L_{1}(j \omega^{\star}) < L_{0}(j \omega^{\star}) \iff G_{c_{1}}(j \omega^{\star}) < G_{c_{0}}(j \omega^{\star})$ for a given $\omega^{\star}$. If $||G_{c_{i}}(s)||_{\infty} = \delta_{i}$, the following can be construed $\forall \omega$.
\begin{eqnarray}
	\begin{split}
		||L_{h}(s)||_{\infty} &= ||G_{c_{i}}(s)||_{\infty} \frac{J_{s} + J_{arm}}{J_{ref}} < 1 \\
		\implies J_{ref}      &> (J_{s} + J_{arm}) \delta_{i}
	\end{split}
\label{eq:OuterLoopGain2}
\end{eqnarray}

According to Equation~\eqref{eq:OuterLoopGain2}, the lower bound on $J_{ref}$ can be decreased if only $\delta_{i}$ is small. Then, the dependency on $J_{arm}$ would also be lower. To execute a small $J_{ref}$, for realizing high frequency position reference, a small inner loop gain, $L_{i}(j \omega)$, at high frequencies is required. This is a general remark from the above analysis; thus emphasizing the importance of the inner loop gain shaping from the \textit{haptic} feedback perspective.

The findings can be summarized as following: (a) a negative torque feedback improves position tracking without compromising the robustness to some extent; (b) a static torque feedback gain might be insufficient to mitigate an overshoot; (c) at least $2^{nd}$-order torque feedback is necessary for robustness increment; and (d) a small loop gain at high frequencies for a fast haptic feedback performance. These are the key takeaways from this section.
\begin{figure}
	\centering
	\includegraphics[width=8.95cm]{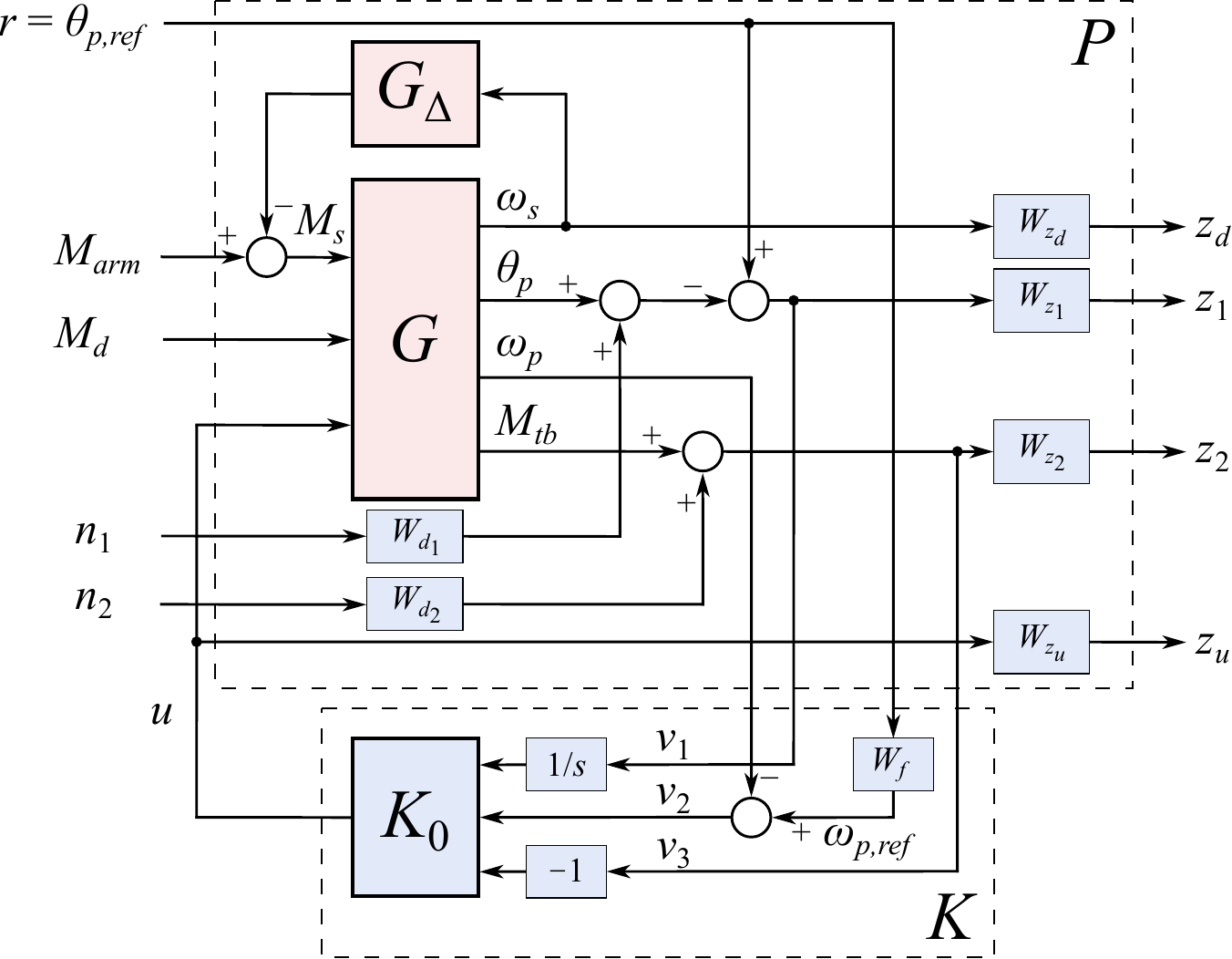}
	\caption{The closed-loop $H_{\infty}$ control configuration with different exogenous input-output channels for the reference position tracking problem. $g_{\Delta}(t) = J_{arm} \dot{\omega}_{s}(t)$ is the inverse additive parametric uncertainty. This $P$-$K$ structure is based on a general formulation, described in~\cite{Skogestad2001}.}
	\label{fig:PK_structure}
\end{figure}

\subsection{$H_{\infty}$ Control Using the LMI Technique}
\label{s:HinfLMI}
The importance of a negative torque feedback in position control using only a static gain is illustrated before. In this section, we aim to design a dynamic torque feedback response based on the LMI$-H_{\infty}$ optimization framework. The two main goals are: (a) a faster tracking and (b) an increased robustness to the arm inertia uncertainty than the two controllers presented previously. The intention is to achieve the latter through a robust linear control design rather than a parameter dependent gain scheduled controller. Hence, no explicit arm inertia estimation is performed and the torque feedback will not change over any parametric uncertainty.
\begin{figure*}
	\centering
	\includegraphics[width=\textwidth]{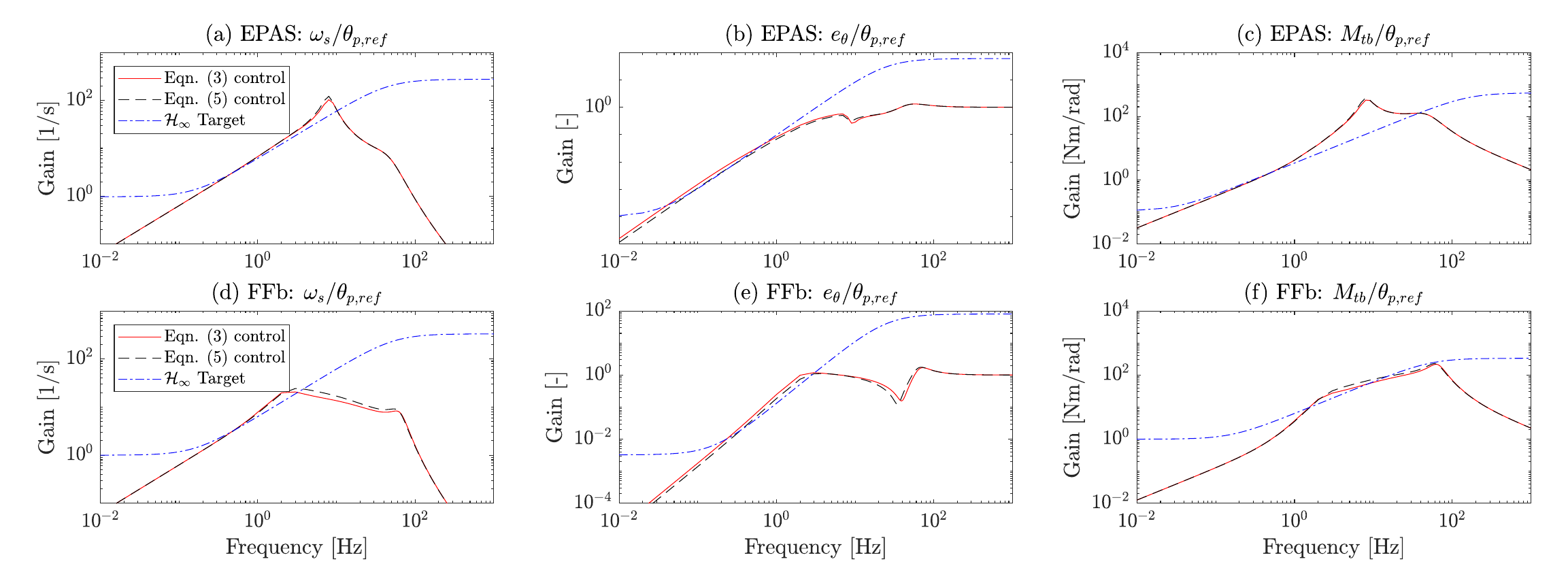}
	\caption{Closed-loop EPAS and FFb frequency response with the position control laws in Equations \eqref{eq:PosCtrlSingleVar} and \eqref{eq:PosCtrlMultiVar}. This formed the basis to design the $H_{\infty}$ targets (and subsequently the respective weighting functions), as shown for $\omega_{s}/\theta_{p,ref}$, $e_{\theta}/\theta_{p,ref}$ and $M_{tb}/\theta_{p,ref}$ respectively.}
	\label{fig:WeightDef}
\end{figure*}

Consider a typical formulation, as shown in Fig.~\ref{fig:PK_structure} for the $P$-$K$ structure, with the generalized plant, $P$, and the feedback controller, $K$. The plant inputs are the exogenous signal $w$: that includes the disturbances, $d(t) = [M_{arm} \ M_{d}]^{T}$, the reference, $r(t) = \theta_{p,ref}$ and the noise, $n(t) = [n_{1} \ n_{2}]^{T}$; and the control signal $u(t)$. The plant outputs are the regulated signals $z(t)$, given by: $z_{d} = W_{z_{d}} \omega_{s}$, $z_{1} = W_{z_{1}} e_{\theta}$, $z_{2} = W_{z_{2}} (M_{tb} + W_{d_{2}} n_{2})$ and $z_{u} = W_{z_{u}} u$. The $H_{\infty}$ norm is obtained by solving the Bounded Real Lemma, see~\cite{Skogestad2001} and~\cite{Gahinet1994} for details. Also, the eigenvalue placement constraint was imposed for real-time compatibility. The implementation was done in the MATLAB Robust Control Toolbox to obtain $K_{0}(s)$. 

In Fig.~\ref{fig:PK_structure}, the problem is formulated as $\theta_{p,ref}$ tracking; which is an interpretation of a typical $S/KS$ mixed-sensitivity optimization~\cite{Skogestad2001}. The disturbances are considered null, i.e. $d = 0$, and $w = [n_{1} \ n_{2} \ r]^{T}$. The generalized plant becomes: 
\begin{eqnarray}
	\begin{split}
		P(s) = \left[
		\begin{array}{ccc|c}
							   0 & 					 0 & 		       0 &  W_{z_{d}} G_{13} \\
			-W_{z_{1}} W_{d_{1}} & 			         0 &       W_{z_{1}} & -W_{z_{1}} G_{23} \\
					           0 & W_{z_{2}} W_{d_{2}} &               0 &  W_{z_{2}} G_{43} \\
							   0 &                   0 &               0 &             W_{u} \\
			\hline
					  -W_{d_{1}} &                   0 &     		   1 & 			 -G_{23} \\
							   0 &                   0 &           W_{f} &           -G_{33} \\
							   0 &           W_{d_{2}} &               0 &            G_{43}
		\end{array}
		\right],
	\end{split}
	\label{eq:GeneralizedPlantFcn}
\end{eqnarray}
such that the controller inputs are $\nu(t) = [e_{\theta} \ \dot{e}_{\theta} \ M_{tb} + d_{2}]^{T}$ and $d_{2} = W_{d_{2}} n_{2}$. We will now explain the motivation behind each weighting function, discuss how they have been derived and the final setup for the closed-loop interconnection.

The performance weights are derived using the prior knowledge of the closed-loop formulation with the single- and the multi-variable control laws in Equation~\eqref{eq:PosCtrlSingleVar} and \eqref{eq:PosCtrlMultiVar} respectively. The corresponding frequency response can be seen in Fig.~\ref{fig:WeightDef}. Equation~\eqref{eq:ClosedLoopTFs} is derived using Section~\ref{s:MultVarPosCtrl} definitions: $L_{i}(s) = K_{\theta}(s) G_{23} \Delta(s)$ and $G_{c}(s) = L_{i}(s)/(1+L_{i}(s))$.
\begin{eqnarray}
	\begin{split}
		\begin{bmatrix}
			\omega_{s}(s) \\
			e_{\theta}(s) \\
			M_{tb}(s)     \\
			M_{mot}(s)
		\end{bmatrix}
		= \frac{G_{c}}{G_{23}}
		\begin{bmatrix}
			G_{13}                   			   \\
			\big(\Delta(s) K_{\theta}(s)\big)^{-1} \\
			G_{43}                  			   \\
			1
		\end{bmatrix}
		\theta_{p,ref}(s)
	\end{split}
	\label{eq:ClosedLoopTFs}
\end{eqnarray}
For the control law in Equation~\eqref{eq:PosCtrlSingleVar}, $K_{M}(s) = 0$ and $\Delta(s) = 1$. We want to achieve, at least, a similar response to the previous controllers. Therefore, the $H_{\infty}$-targets have been designed in a close proximity to the existing controllers' FRF response in the relevant frequency range for each variable and with the same order, see Fig.~\ref{fig:WeightDef}. The target function's cut-off frequency is set to $25$~Hz. These transfer functions are biproper and with no simple poles; which is necessary for LMI feasibility. $W_{z_{d}}$, $W_{z_{1}}$, $W_{z_{2}}$ and $W_{z_{u}}$ are subsequently inverse of their targets. $W_{z_{u}}$ is excluded in Fig.~\ref{fig:WeightDef} due to space constraints, but it follows the same principle. See Appendix~\ref{app:Weights} for the values.

To mimic the reality closer, we would need some additive noise to the measured channels: $\theta_{p}$ and $M_{tb}$. Subsequently, we use $1^{\text{st}}$-order high pass filters for the signal noise captured from the standalone sensor measurements, see Appendix~\ref{app:Weights}. The signal offsets were corrected beforehand; therefore we exclude any dynamic free terms.

The next discussion point is on the $H_{\infty}$ loop shaping procedure. In Fig.~\ref{fig:PK_structure}, an integrator on $\nu_{1}$ is included. This ensures low steady state tracking error (defined by $e_{\theta}$) and good disturbance rejection at low frequencies. Without this integrator, $K_{0}(s)$ synthesis would be biproper; hence $|e_{\theta}(j \omega)/r(j \omega)| > 0$, as $\omega \to 0$. Therefore, with an enforced integral state on $\nu_{1}$, we ensure a strictly proper control law with relative degree $1$, s.t.
\begin{eqnarray}
	\begin{split}
		\lim\limits_{\omega \rightarrow 0} \bigg|\frac{u (j \omega)}{e_{\theta} (j \omega)}\bigg| = \infty \text{ and } \lim\limits_{\omega \rightarrow \infty} \bigg|\frac{u (j \omega)}{e_{\theta} (j \omega)}\bigg| = 0.
	\end{split}
	\label{Reform_HinfLMI}
\end{eqnarray}

Due to the integral action on the error, $e_{\theta}$, there is a minimal control actuation for high frequencies. As a result, the tracking performance is compromised. To avoid this problem, $\nu_{2} = \dot{e}_{\theta}$ is explicitly used as a separate input channel, which guarantees the following: $|u (j \omega)/e_{\theta} (j \omega)| > 0$, as $\omega \to \infty$. For computing $\dot{e}_{\theta}$, we require $\omega_{p,ref}$. Therefore, a $1^{\text{st}}$-order high pass filter, $W_{f}(s) := [A_{f}, B_{f}, C_{f}, D_{f}]$, is implemented with $15$~Hz cut-off frequency. If $\dot{e}_{\theta}$ is excluded, then the high frequency tracking would be poor due to the integral state of the controller.
\begin{figure*}
	\centering
	\includegraphics[width=\textwidth]{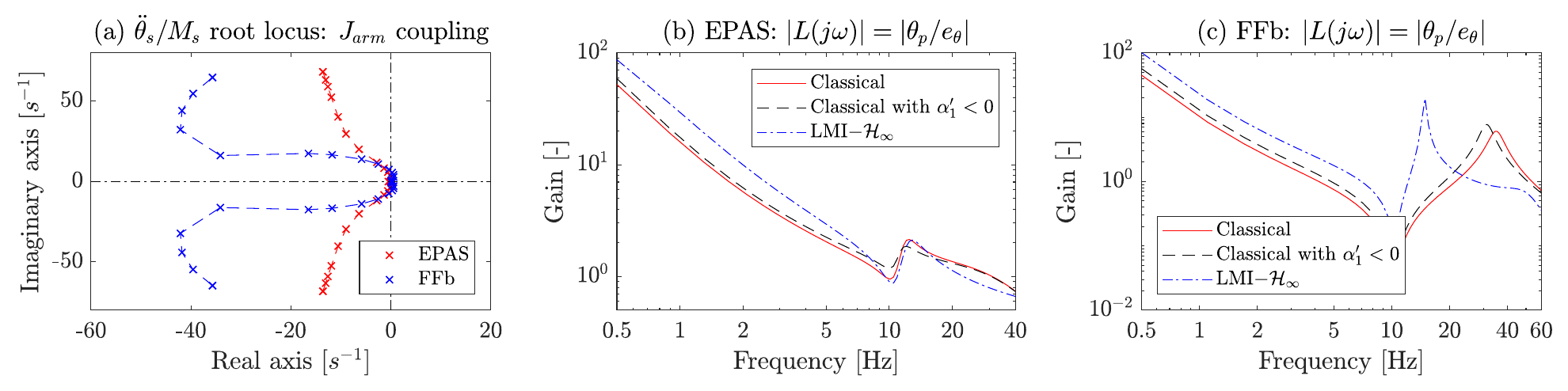}
	\caption{(a) EPAS and FFb root locus plot for the $H_{\infty}$ controlled driver interaction port, $\dot{\omega}_{s}/M_{s}$, with an increasing coupling inertia, $J_{arm} \in [0,\infty)$. The driver uncoupled loop gain for (b) EPAS and (c) FFb with the control laws in Equations \eqref{eq:PosCtrlSingleVar}, \eqref{eq:PosCtrlMultiVar} and $K(s)$ from Fig.~\ref{fig:PK_structure} respectively.}
	\label{fig:RobustStability}
\end{figure*}

$\nu_{3}(t) = M_{tb}(t) + d_{2}(t)$ is multiplied with $-1$ to ensure a negative torque feedback, following the result in Section~\ref{s:MultVarPosCtrl}.

\textit{Remark 3.1:} For a real-time digital control application, we are bounded by the smallest possible discrete time step of the computing machine. Therefore, we require another inequality condition to ensure that the continuous-time controller $K_{0}(s)$ is causal and implementable in a real-time machine.

For numerical stability with the explicit Euler solver, we must constrain the eigenvalues of $K_{0}(s)$ in the circular disc $\mathcal{D}:=|1 + h \lambda_{i}| \leq 1$, $\forall \ \lambda_{i} \in \mathbb{C}$, where $\lambda_{i}$ are the eigenvalues and $h = 1$ ms is the solver time step. Constraining the eigenvalues of $A_{k}$ does not represent a convex LMI problem. Instead, we constrain the closed-loop eigenvalues of $A_{cl}$ to $\mathcal{D}$, given that the no large real part eigenvalue of the open loop plant $A$. Then, the sufficient conditions in~\cite{Chilali1996} ensure the eigenvalues of $A_{k}$ are in $\mathcal{D}$, see~\cite{Gahinet1994} for $A$, $A_{k}$ and $A_{cl}$ definitions.

To decrease the computational effort of a real-time controller, the model order of $K_{0}(s)$ is reduced from $13$ to $5$ by following the balanced residualization, see~\cite{Skogestad2001}; thus removing the fast eigenmodes. As a consequence, the $H_{\infty}$ norm of the closed-loop tracking error, $||G_{c}(s) - G_{c}^{a}(s)||_{\infty}$, is $0.0454$ (at $9.82$~Hz for EPAS) and $0.0807$ (at $2.86$~Hz for FFb) with $J_{arm} = 0.03$ kgm$^{2}$, where $G_{c}^{a}(s)$ is the approximated system.

\textit{Remark 3.2:} Given an uncertainty in $J_{arm}$, the closed-loop system should remain stable with the obtained $K(s)$, for all the plant models with $J_{arm} \in [0,\infty)$, to ensure robust stability~\cite{Briat2015, Skogestad2001}. To show this, consider $p = 1/(J_{s} + J_{arm})$ as a linear time varying parameter in $A(p)$, s.t. $p \in [0, 1/J_{s}] := \mathcal{P}$. 

Computing the controller, $K(s)$ in Fig.~\ref{fig:PK_structure}, using the $H_{\infty}$ solution, $K_{0} = [A_{k}, B_{k}, C_{k}, D_{k}]$, the input signals: $r = \theta_{p,ref}$ as the reference and $y = [\theta_{p} \ \omega_{p} \ M_{tb}]^{T}$ as the output feedback; and including the loop shaping functions. Then, the closed-loop interconnection is formed using the plant in Equation~\eqref{eq:StateSpace} and $K(s)$. Hence, the stability is defined by $A_{cl}'(p) =$
\begin{eqnarray}
	\begin{split}
		\begin{bmatrix}
			A(p) - B_{2} D_{k_{2}}' C & B_{2} C_{k} & 0 & B_{2} D_{k_{1}} & B_{2} D_{k_{2}} C_{f} \\
			-B_{k_{2}}' C & A_{k} & 0 & B_{k_{1}} & B_{k_{2}} C_{f} \\
			     0 & 0 & 0 & 1 & 0    \\
			-C_{1} & 0 & 0 & 0 & 0    \\
			     0 & 0 & 0 & 0 & A_{f}
		\end{bmatrix},
	\end{split}
	\label{eq:LyapStabEqn}
\end{eqnarray}
s.t. $D_{k_{2}}' C = D_{k_{2}} C_{2} + D_{k_{3}} C_{3}$ and $B_{k_{2}}' C = B_{k_{2}} C_{2} + B_{k_{3}} C_{3}$.

Stability can be checked by the eigenvalues of $A_{cl}'(p)$. The root locus plot of $\dot{\omega}_{s}/M_{s}$, in Fig.~\ref{fig:RobustStability}(a), is used to ensure closed-loop stability of all the plant models within $\mathcal{P}$. The two dominant conjugate eigenvalues of $A_{cl}'(p)$ (corresponding to the steering wheel's equation of motion) always remain in the left half plane; and gradually converge towards the origin, such that $(\theta_{s}, \omega_{s}) = (0,0)$ as $J_{arm} \to \infty$. Thus, the derived $H_{\infty}$ controller ensures robust stability.

\textit{Remark 3.3:} The loop gain of different controllers: classical ($L_{0}(s)$); classical with a negative torque feedback ($L_{1}(s)$); and $H_{\infty}$ ($L_{1,H_{\infty}}(s)$) are shown in Fig.~\ref{fig:RobustStability}(b) for EPAS and in Fig.~\ref{fig:RobustStability}(c) for FFb, respectively. They are derived for uncoupled driver and environment ports with $J_{arm} = 0$ kgm$^2$. From these figures, it can be concluded that we have achieved a desired loop gain shape using the LMI-$H_{\infty}$ framework, i.e., a larger loop gain at low frequencies and a smaller loop gain at high frequencies.

\section{Results}
\label{s:Results}
A comparison of proposed real-time digital controllers using the closed-loop measurements is covered in this section.

\subsection{Experimental setup}
\label{s:RT_setup}
\begin{figure}
	\centering
	\includegraphics[width=8.65cm]{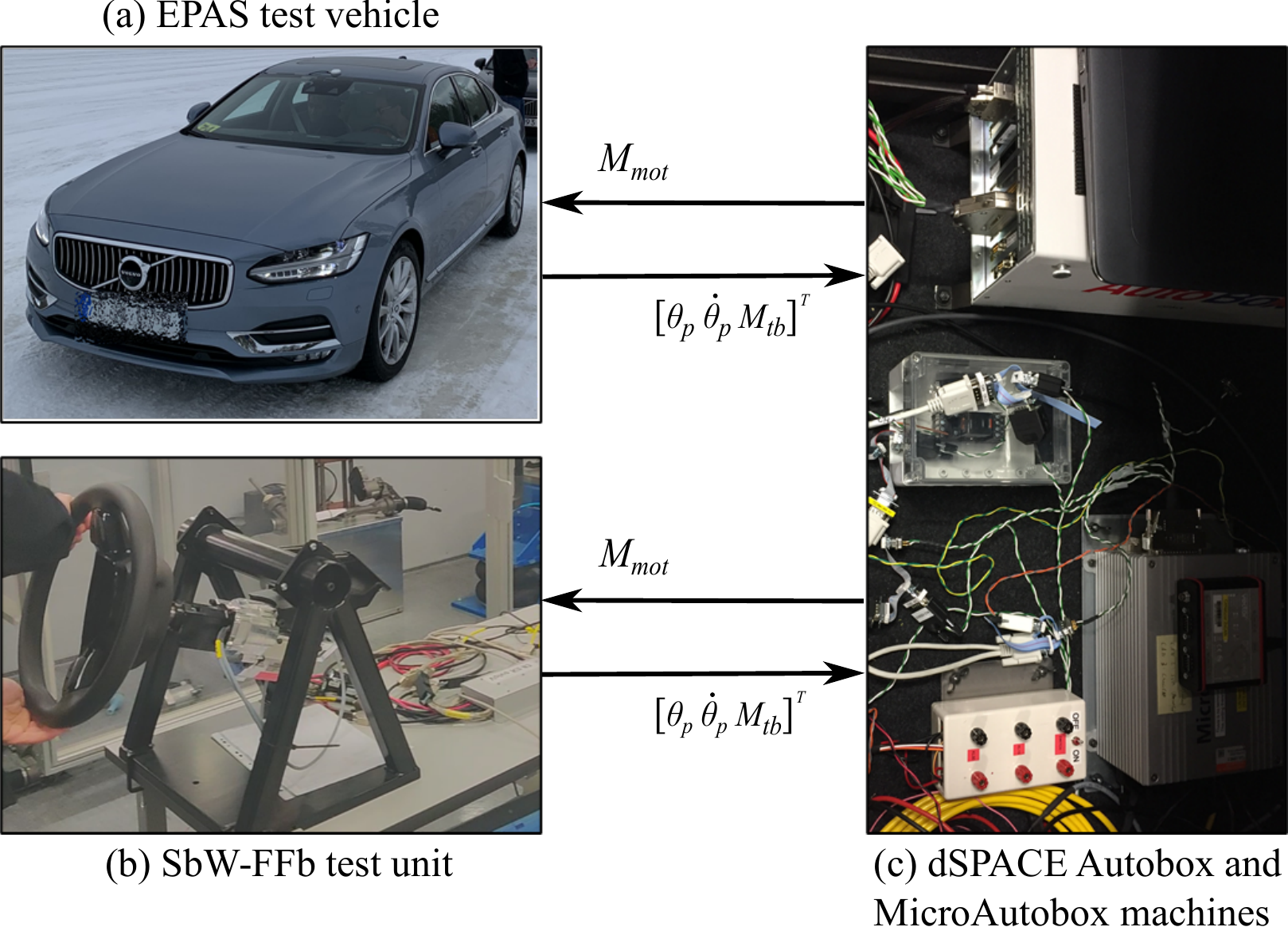}
	\caption{The experimental setup for (a) EPAS and (b) FFb systems.}
	\label{fig:ExpSetup}
\end{figure}

The EPAS experiments were performed on a Volvo S90 test vehicle equipped with a dSPACE Autobox DS$1007$. Whereas for the SbW-FFb experiments, the test rig was equipped with a MicroAutobox $1401$, see Fig.~\ref{fig:ExpSetup}. The sensed input signals were pinion angle, pinion speed, and torsion bar torque; the output signal was motor torque request. They were interfaced using a private controller area network (CAN) bus at $1$~kHz sampling frequency. Therefore, the continuous-time LMI$-H_{\infty}$ state-space matrices were discretized at $1$ ms using the explicit Euler method. The execution was performed using the real-time interface blocks of the processor board in MATLAB/Simulink. The FRF measured signals were post-processed using a zero-phase $30$~Hz low-pass 4th-order Butterworth filter. Moreover, the Coulomb motor friction torque was compensated using a state estimator. The estimated motor friction torque, $\hat{M}_{mot,fric}$, was added to the controller request, such that $M_{mot,tot}(t) = M_{mot}(t) + \hat{M}_{mot,fric}(t)$. For a given system, the same friction compensation was used with different controllers.

\subsection{Frequency Response Comparison}
\label{s:FRF_Comp}
In this comparison, we have chosen to illustrate the tracking performance using $\theta_{p}/\theta_{p,ref}$ FRF plots. A sine sweep signal was used for the reference excitation with a maximum amplitude of $2^\circ$ from $0.01-8$~Hz for EPAS and $0.01-15$~Hz for FFb systems. A higher $\theta_{p,ref}$ excitation request for high inertial systems at such frequencies should be avoided due to the saturation limits of the torque sensor and the applied motor torque. This would defy the purpose of linear robust control.
\begin{figure*}
	\centering
	\includegraphics[width=16.15cm]{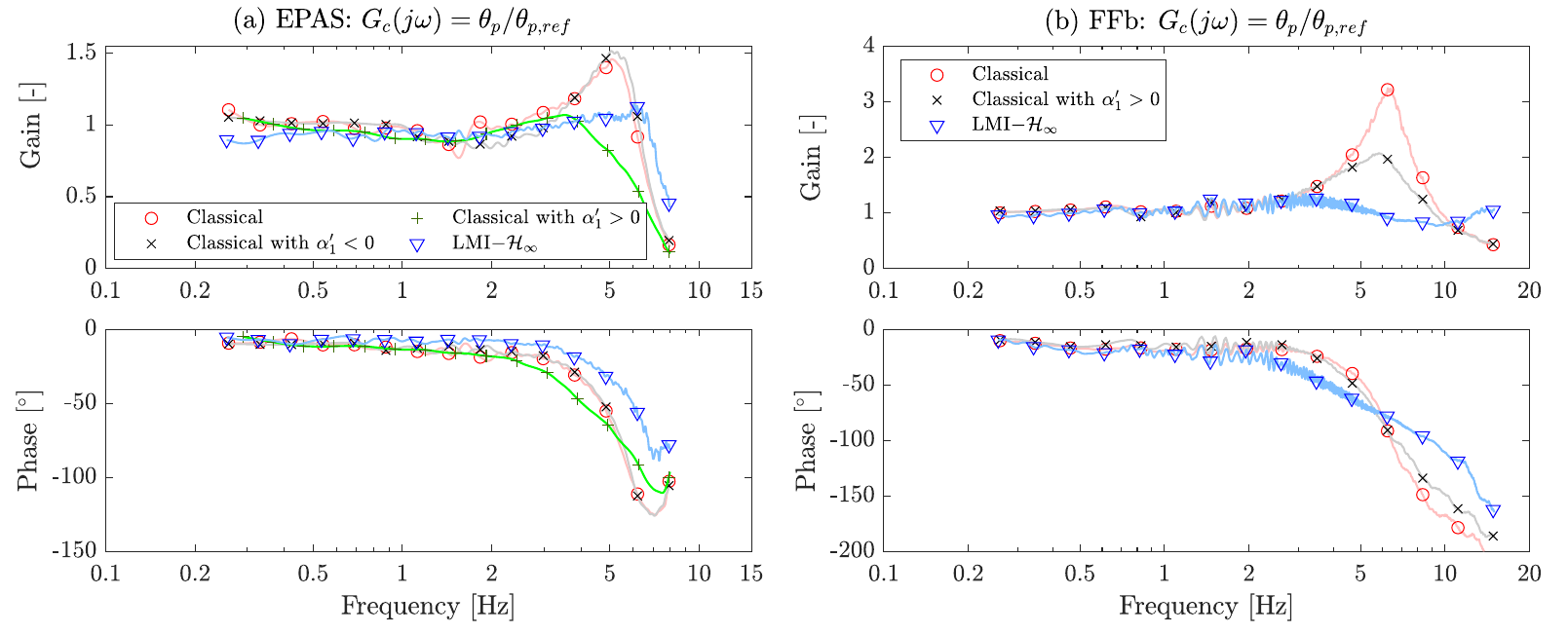}
	\caption{Measured FRF plots from experiments in: (a) EPAS test vehicle and (b) FFb hardware. The comparison shows the tracking performance, $\theta_{p}/\theta_{p,ref}$, for different position controllers: classical ($G_{c_{0}}(s)$), classical with a static torque feedback ($G_{c_{1}}(s)$) and $H_{\infty}$ ($G_{c_{1,H_{\infty}}}(s)$).}
	\label{fig:FRF_Result}
\end{figure*}

Comparing different EPAS controllers, we present the FRF plots in Fig.\ref{fig:FRF_Result}(a). The classical performance given by $G_{c_{0}}(s)$ shows a response similar to a typical $2^{\text{nd}}$-order transfer function. The cut-off frequency is approximately $6.35$~Hz with a low damping ratio. The classical controller has been tuned to meet the desired objectives of vehicle motion and haptic feedback control performance. The nominal overshoot can be reduced by either decreasing $\beta_{0}$ or increasing $\beta_{3}$, but this would affect the low frequency tracking error i.e. higher $e_{\theta}$ or $\dot{\omega}_{s}$ noise amplification in the steering response respectively. 

With $\alpha_{1}' < 0$, $G_{c_{1}}(s)$ improves marginally at the expense of a slightly higher overshoot, hence upholding \textit{Proposition 3.1} and \textit{3.2}. The respective controller gains are given in Table~\ref{t:ControlGains}. For understanding, we have also provided $G_{c_{1}}(s)$ with a positive torque feedback, i.e. $\alpha_{1}' = 0.175$, and the same $K_{\theta}(s)$. As anticipated, this controller response is the slowest and most damped among others. Last but not the least, $G_{c_{1,H_{\infty}}}(s)$ relatively outperforms the other controllers. The cut-off frequency is $7.75$~Hz with an overshoot reduction. Although the gain response is more damped but the improvement can be seen in phase delay at high frequencies, resulting in a faster response of the $H_{\infty}$ controller. For low frequency tracking, i.e., less than $1$~Hz, the gain is comparable and $G_{c_{1,H_{\infty}}}(s)$ has lower delay.
%\begin{figure}
%	\centering
%	\includegraphics[width=8.165cm]{Figures/Fig_FricFbLin_Result_v2p6.pdf}
%	\caption{Closed-loop FFb performance using the classical position controller: (a) without and (b) with the friction compensation respectively.}
%	\label{fig:FbLin_Result}
%\end{figure}

The main reason for not achieving a significant bandwidth increase for the $H_{\infty}$ controller is caused by low resolution of $\theta_{p}$ and $\omega_{p}$ signals. This affects enormously for a small $\theta_{p,ref}$ excitation. The difference is more prominent at large reference values but then we are more likely to end up in the saturation limit of either the torque sensor or the actuator. Within the linear operational domain, the difference in response between $G_{c_{1,H_{\infty}}}(s)$ and other transfer functions is subtle but noticeable.

We will now look at the FRF plots for SbW-FFb shown in Fig.\ref{fig:FRF_Result}(b). $G_{c_{0}}(s)$ represents the classical controller's performance. The cut-off frequency is approximately $11.25$~Hz with low damping ratio, thus causing a high overshoot, which is undesirable. The attributing reasons were high quantization noise and low resolution of $\omega_{p}$ signal. Subsequently, these effects were mitigated by using a $15$~Hz $1^{\text{st}}$-order low pass filter. As a result, we were restricted to low $\beta_{2}$ value; hence, an inferior performance in comparison to EPAS in terms of overshoot and delay. To avoid an even higher overshoot, $\alpha_{1}' < 0$ iteration is omitted. With $\alpha_{1}' > 0$ and the classical $K_{\theta}(s)$ from Table~\ref{t:ControlGains}, the low frequency loop gain, tracking performance, and overshoot can be reduced, according to \textit{Proposition 3.1}.
\begin{figure*}
	\centering
	\includegraphics[width=17.25cm]{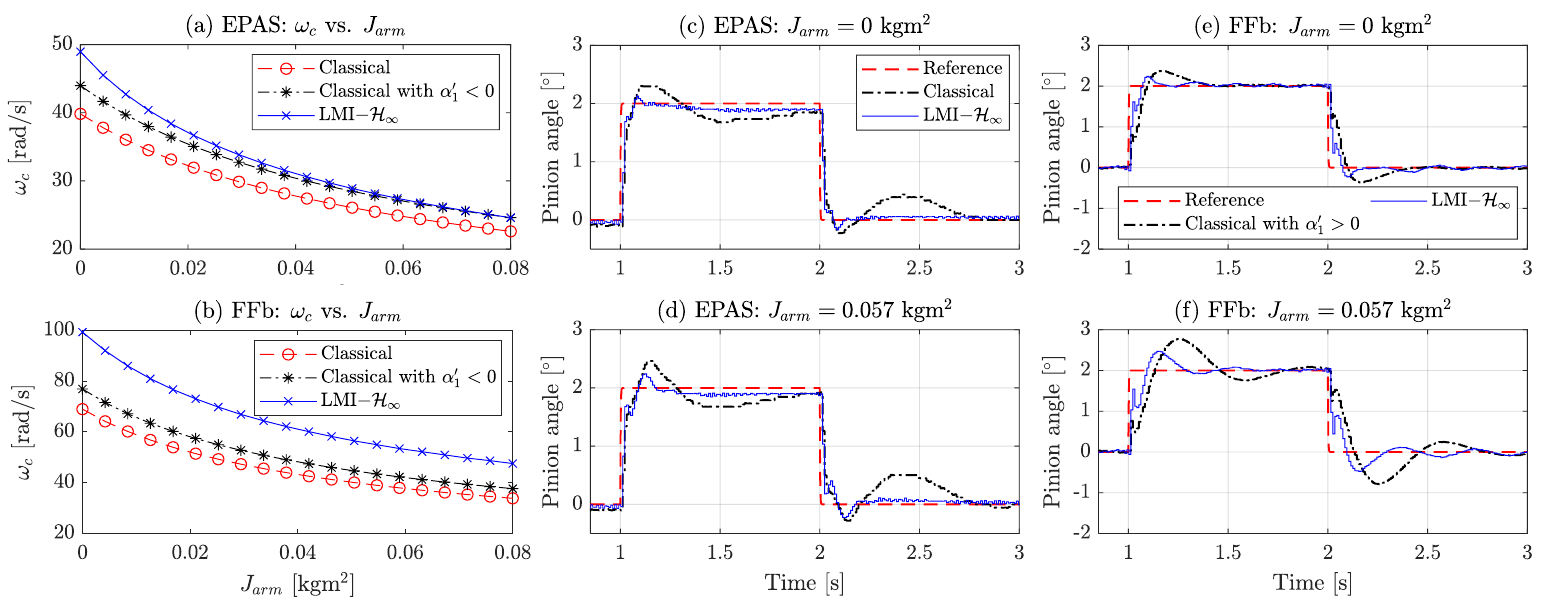}
	\caption{The tracking cut-off frequency over $J_{arm}$ uncertainty in (a) EPAS and (b) FFb for different position controllers: classical, classical with a static torque feedback, and $H_{\infty}$. The closed-loop step response with different controllers and arm inertia values in EPAS, (c) and (d), and FFb, (e) and (f), respectively.}
	\label{fig:Robustness_Result}
\end{figure*}

A better FFb tracking performance is given by $G_{c_{1,H_{\infty}}}(s)$ in Fig.~\ref{fig:FRF_Result}(b), with a higher cut-off frequency than $G_{c_{0}}(s)$ and $G_{c_{1}}(s)$. Due to high damping, $|G_{c_{1,H_{\infty}}}(s)|$ is close to $1$ for a large frequency interval; and a low overshoot within $2$-$6$~Hz, but with a higher phase delay. Here, the main contribution of torque feedback was to reduce the nominal overshoot, as seen in $G_{c_{0}}(s)$, and a faster response at high frequencies than others. Whereas, the low frequency tracking is comparable. Although the frequency response measurements were performed with $J_{arm} = 0$ kgm$^2$, but we have shown an improvement with an LMI$-H_{\infty}$ controller regardless of the system. Hence, with an optimal torque feedback, better tracking and lower phase/time delay are achievable than the classical solution.

\subsection{Performance and Robustness Under Uncertainty}
\label{s:Uncertainty_Comp}
This section is devoted to investigate the effect of torque feedback with respect to the coupled arm inertia uncertainty.

Performance and robustness can be adjudged in terms of cut-off frequency, $\omega_{c}$, and its change over the uncertainty in $J_{arm}$. The corresponding $\omega_{c}$ values are shown in Fig.~\ref{fig:Robustness_Result}(a) and (b) for different controllers: classical, classical with a static torque feedback ($\alpha_{1}' < 0$) and LMI$-H_{\infty}$.

For EPAS, $\omega_{c_{0}}$ decreases with an increasing $J_{arm}$. The SISO phase margin, $\phi_{m_{0}}$, for $L_{0}(s)$ is $60^\circ$ and it remains unaffected. Using $\alpha_{1}'$ from Table~\ref{t:ControlGains}, $\omega_{c_{1}}$ shows a similar behavior as $\omega_{c_{0}}$, with $10\%$ higher performance in Fig.~\ref{fig:Robustness_Result}(a). However, $G_{c_{1,H_{\infty}}}(s)$ performs the best at low $J_{arm}$ values due to a large loop gain, $L_{1,H_{\infty}}(s)$, at low frequencies. Consequently, $\omega_{c_{1,H_{\infty}}}$ is roughly $20\%$ higher with respect to $\omega_{c_{0}}$ at zero inertia. At high $J_{arm}$, $\omega_{c_{1,H_{\infty}}} > \omega_{c_{0}}$ is still maintained but to a lesser extent.

The FFb results are shown in Fig.~\ref{fig:Robustness_Result}(b). As $J_{arm}$ increases, $\omega_{c_{0}}$ decreases. Also, $\phi_{m_{0}}$ for $L_{0}(s)$ decreases from $40^\circ$ to $30^\circ$. $\phi_{m_{0}}$ variation primarily differs between EPAS and FFb because of $J_{p}$. Due to higher motor-to-rack gear ratio and wheel unsprung mass in EPAS, $J_{p} \gg J_{arm}$, whereas $J_{p} < J_{arm}$ for FFb. With $\alpha_{1}' < 0$, $\omega_{c_{1}}$ can be marginally increased as compared to $\omega_{c_{0}}$ for a given $J_{arm}$, see Fig.~\ref{fig:Robustness_Result}(b). For the $H_\infty$ controller, $G_{c_{1,H_{\infty}}}(s)$ shows a substantial performance and robustness improvement. Hence, $\omega_{c_{1,H_{\infty}}} > \omega_{c_{0}}$ by roughly $42\%$, regardless of $J_{arm}$. However, $\omega_{c_{1,H_{\infty}}}$ also decreases with an increasing $J_{arm}$ due to the causality of position control.

Let us compare the simulation results of different controllers for a step response in Fig.~\ref{fig:Robustness_Result}(c)-(f). This is performed for an uncoupled driver port with and without a certain arm inertia. $J_{arm} = 0.057$ kgm$^2$ was selected based on the estimated value from~\cite{Schenk2021}. Here, we illustrate a simulated response rather than experiments to avoid the influence of other physical impedance parameters, i.e., stiffness and damping. In both EPAS and FFb, for a given $J_{arm}$, the $H_{\infty}$ controller has relatively faster response, lower overshoot and lower settling time as compared to the classical controllers; thus more robust.

With $J_{arm} = 0.057$ kgm$^2$, the respective responses exhibit a higher overshoot and a longer settling time in comparison to zero inertia. But the relative performance of the $H_{\infty}$ controller is still better. Thus, it illustrates an improved robustness under $J_{arm}$ uncertainty. For the FFb results in Fig.~\ref{fig:Robustness_Result}(e)-(f), we have disregarded the classical controller due to very low damping ratio. Thereby, the comparison is done with the classical controller including a static positive torque feedback gain.

\section{Conclusion}
\label{s:Conclusion}
This paper has investigated a robust position/angle controller for electric power assisted steering and steer-by-wire force-feedback systems based on the LMI$-H_{\infty}$ optimization framework. Such a controller is typically required for controlling: (a) the vehicle's lateral motion (especially when the driver is hands-off) and (b) the haptic feedback reference, i.e., when the driver is in the loop. Therefore, the proposed solution ensures robustness against the driver's coupled arm inertia.

A classical position controller includes angular position and velocity as the feedback states. Whereas, we present a multi-variable feedback solution by also adding the sensed torque information. A static negative torque feedback gain can improve reference tracking but at the expense of relatively higher overshoot and similar robustness. Therefore, an \textit{optimal} torque feedback response is synthesized for a desired loop gain shape, i.e., a large loop gain at low frequencies for tracking performance and a small loop gain at high frequencies for robustness and to realize high frequency haptic feedback position reference. The experiments illustrated an improvement in reference tracking and robustness with the proposed position controller against the coupled arm inertia. However, a decreasing performance with an increasing coupled inertia still persist because of the causality. 

In comparison to a torque controller, the position reference tracking requires further improvement under high coupled inertia. Though the proposed LMI-$H_{\infty}$ controller is robust for vehicle motion control, it should be further improved for high frequency steering feedback reference. Future work will extend the current research with more advanced $H_{\infty}$ concepts such as $2$-DOF, linear scheduling with arm inertia, and $\mu$-synthesis.

\appendices
%Appendixes, if needed, appear before the acknowledgment.

\section{List of Variables and Transfer Functions}
\label{app:Definition}
%A list of variables and transfer functions' symbols and the respective descriptions are given in Table~\ref{t:Definition}.
The description is given in Table~\ref{t:Definition}.
\begin{table}
	\centering
	\caption{Description of Variables and Transfer Functions}
	\setlength{\tabcolsep}{3pt}
	\begin{tabular}{|p{28pt}|p{70pt}|p{40pt}|p{87pt}|}
		\hline
		Symbol & Description & Symbol & Description \\
		\hline
		$e_{\theta}, \dot{e}_{\theta}$ & Position error states & $L_{i}(s)$      & Inner loop gain  		  \\
		$\omega_{c}$ 				   & Controller bandwidth  & $L_{h}(s)$      & Overall loop gain          \\
		$\omega_{\phi_{m}}$            & Gain crossover freq.  & $L_{0}(s)$ 	 & Classical loop gain        \\
		$G_{ij}(s)$ 			       & SISO open loop plant  & $L_{1}(s)$ 	 & Multi-variable loop gain   \\
		$K_{\theta}(s)$ 			   & Position control law  & $L_{1,H_{\infty}}(s)$ & $H_{\infty}$ loop gain \\
		$K_{M}(s)$      			   & Torque control law    & $G_{c_{0}}(s)$  & Classical closed-loop      \\
		$\Delta(s)$     			   & Torque feedback loop  & $G_{c_{1}}(s)$  & Multi-variable closed-loop \\
		$H_{ref}(s)$                   & Haptic reference      & $G_{c_{1,H_{\infty}}}(s)$ & $H_{\infty}$ closed-loop \\
		\hline
		\multicolumn{4}{p{225pt}}{}
	\end{tabular}
	\label{t:Definition}
\end{table}

\section{Uncoupled Torque Feedback Transfer Function}
\label{app:LoopGain}
The uncoupled torque feedback transfer function, $\Delta(s)$, for the loop gain, $L(s) = K_{\theta}(s) G_{23} \Delta(s)$, is as follows
\begin{eqnarray}
	\begin{split}
		\Delta(s)    &= \frac{b_{3} s^3 + b_{2} s^2 + b_{1} s + b_{0}}{a_{3} s^3 + a_{2} s^2 + a_{1} s + a_{0}} %\\
%		\noindent \text{where} &, \\
%		a_{3}        &= b_{3} =  J_{s} J_{p} \\
%		a_{2}        &= b_{2} =  J_{s} b_{p} + J_{p} b_{s} \\
%		a_{1}        &= (J_{s} (1 + \alpha_{1}' i_{mot}) + J_{p}) c_{tb} + b_{s} b_{p} \\
%		b_{1}        &= (J_{s} + J_{p}) c_{tb} + b_{s} b_{p} \\
%		a_{0}        &= c_{tb} (b_{s} (1 + \alpha_{1}' i_{mot}) + b_{p}) \\
%		b_{0}        &= c_{tb} (b_{s} + b_{p}).
	\end{split}
	\label{eq:UncoupledLoopGain}
\end{eqnarray}
where $a_{3} = b_{3} = J_{s} J_{p}$, $a_{2} = b_{2} = J_{s} b_{p} + J_{p} b_{s}$, $a_{1} = (J_{s} (1 + \alpha_{1}' i_{mot}) + J_{p}) c_{tb} + b_{s} b_{p}$, $b_{1} = (J_{s} + J_{p}) c_{tb} + b_{s} b_{p}$, $a_{0} = c_{tb} (b_{s} (1 + \alpha_{1}' i_{mot}) + b_{p})$ and $b_{0} = c_{tb} (b_{s} + b_{p})$.

\section{$H_{\infty}$ Weighting Functions}
\label{app:Weights}
The weighting functions are as follows: $W_{f}(s) = s/(1 + T_{f} s)$, $W_{z_{d}}(s) = w_{z_{d}} (1 + T_{z_{d}} s)/(1 + T_{p_{d}} s)$, $W_{z_{1}}(s) = w_{z_{1}} (1 + T_{z_{1}} s)^2/(1 + T_{p_{1}} s)^2$, $W_{z_{2}}(s) = -w_{z_{2}} (1 + T_{z_{2}} s)/(1 + T_{p_{2}} s)$, $W_{z_{u}}(s) = w_{z_{u}} (1 + T_{z_{u}} s)/(1 + T_{p_{u}} s)$, $W_{d_{1}}(s) = w_{d_{1}} s/(1 + T_{d_{1}} s)$, and $W_{d_{2}}(s) = w_{d_{2}} s/(1 + T_{d_{2}} s)$.
%\begin{eqnarray}
%	\begin{split}
	%		W_{f}(s)     &= s/(1 + T_{f} s) \\
	%		W_{z_{d}}(s) &= w_{z_{d}} (1 + T_{z_{d}} s)/(1 + T_{p_{d}} s) \\
	%		W_{z_{1}}(s) &= w_{z_{1}} (1 + T_{z_{1}} s)^2/(1 + T_{p_{1}} s)^2 \\
	%		W_{z_{2}}(s) &= -w_{z_{2}} (1 + T_{z_{2}} s)/(1 + T_{p_{2}} s)  \\
	%		W_{z_{u}}(s) &= w_{z_{u}} (1 + T_{z_{u}} s)/(1 + T_{p_{u}} s) \\
	%		W_{d_{1}}(s) &= w_{d_{1}} s/(1 + T_{d_{1}} s) \\
	%		W_{d_{2}}(s) &= w_{d_{2}} s/(1 + T_{d_{2}} s)
	%	\end{split}
%\label{eq:WeightFcns}
%\end{eqnarray}
The corresponding parameters are provided in Table~\ref{t:WeightFcns}.
\begin{table}
	\centering
	\caption{Weighting Function Parameters}
	\setlength{\tabcolsep}{3pt}
	\begin{tabular}{|p{45pt}|p{30pt}|p{30pt}|p{45pt}|p{30pt}|p{30pt}|}
		\hline
		Parameter (units) & 
		Value (EPAS)      &
		Value (FFb)       &
		Parameter (units) & 
		Value (EPAS)      &
		Value (FFb)       \\
		\hline
		$T_{f}$[s]       & $0.0106$ & $0.0106$ & $T_{d_{1}}$[s]   	   & $0.005$  & $0.005$    \\
		$T_{z_{d}}$[s]   & $0.0064$ & $0.003$  & $T_{d_{2}}$[s]   	   & $0.01$   & $0.01$     \\
		$T_{p_{d}}$[s]   & $1$      & $0.50$   & $w_{z_{d}}$[s$^{-1}$] & $1.031$  & $0.50$     \\
		$T_{z_{1}}$[s]   & $0.0064$ & $0.0064$ & $w_{z_{1}}$[$-$]      & $10178$  & $107.95$   \\
		$T_{p_{1}}$[s]   & $5$      & $0.50$   & $w_{z_{2}}$[rad/Nm]   & $9.15$   & $1$        \\
		$T_{z_{2}}$[s]   & $0.0064$ & $0.003$  & $w_{z_{u}}$[rad/Nm]   & $0.30$   & $0.20$     \\
		$T_{p_{2}}$[s]   & $5$      & $1$      & $w_{d_{1}}$[$-$]      & $0.0001$ & $0.0001$   \\
		$T_{z_{u}}$[s]   & $0.002$  & $0.003$  & $w_{d_{2}}$[$-$]      & $0.0002$ & $0.0001$   \\
		$T_{p_{u}}$[s]   & $0.05$   & $0.50$   &                       &          &            \\
		\hline
	\end{tabular}
	\label{t:WeightFcns}
\end{table}

\bibliographystyle{IEEEtran}
\bibliography{Bibliography_HinfLMI}
\vspace{-1cm}
\begin{IEEEbiography}[{\includegraphics[width=1in,height=1.25in,clip,keepaspectratio]{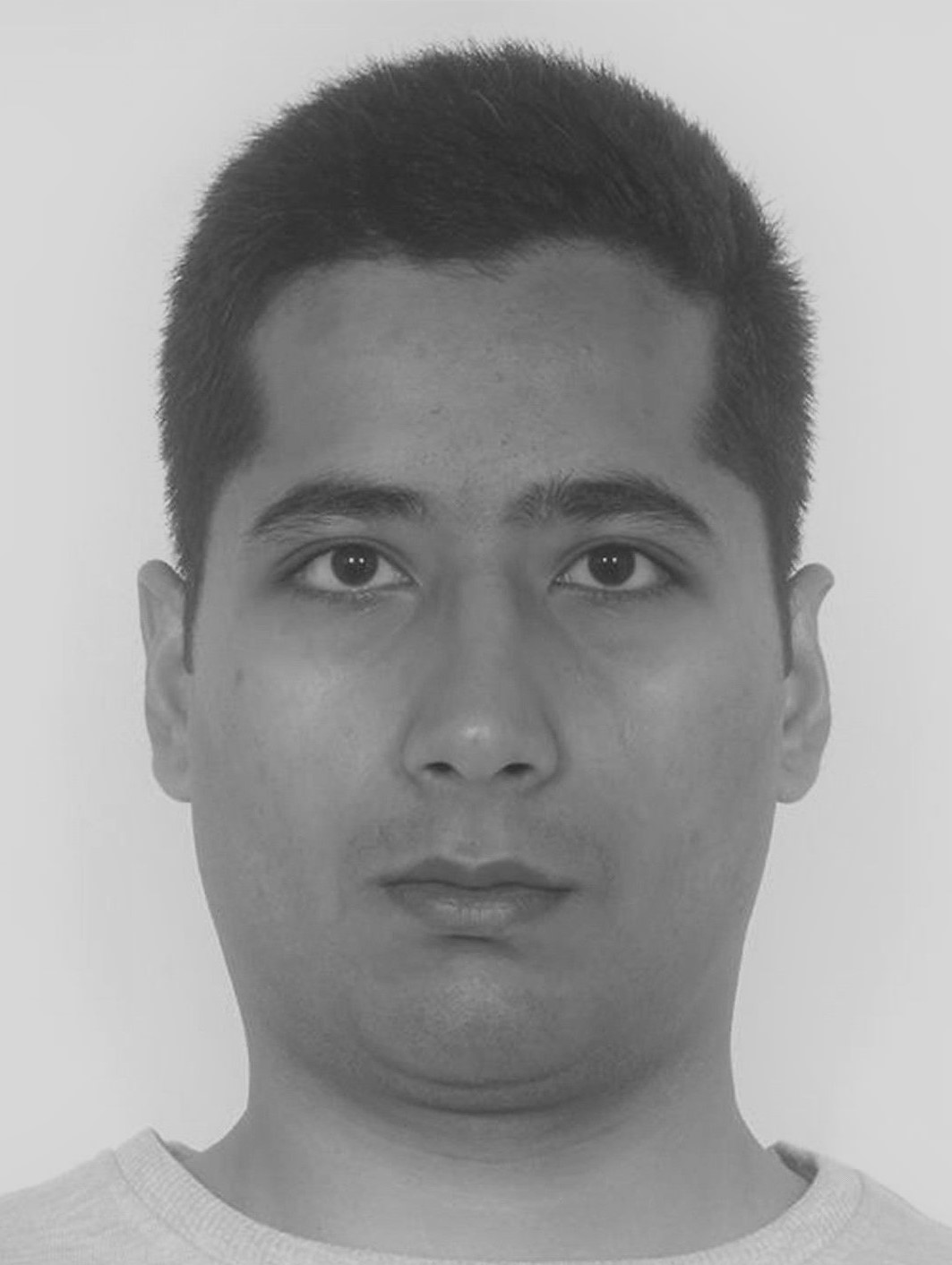}}]{Tushar Chugh}
	received the Ph.D. degree in machine and vehicle systems from Chalmers University of Technology, Gothenburg, Sweden, in 2021 and the M.Sc. degree in automotive engineering from RWTH Aachen University, Aachen, Germany, in 2016. He is currently a senior research engineer with Volvo Car Corporation, Gothenburg, Sweden. His research interests include closed-loop steering control, real-time state estimation, haptic feedback, and vehicle motion control.
\end{IEEEbiography}
\vspace{-1cm}
\begin{IEEEbiography}[{\includegraphics[width=1in,height=1.25in,clip,keepaspectratio]{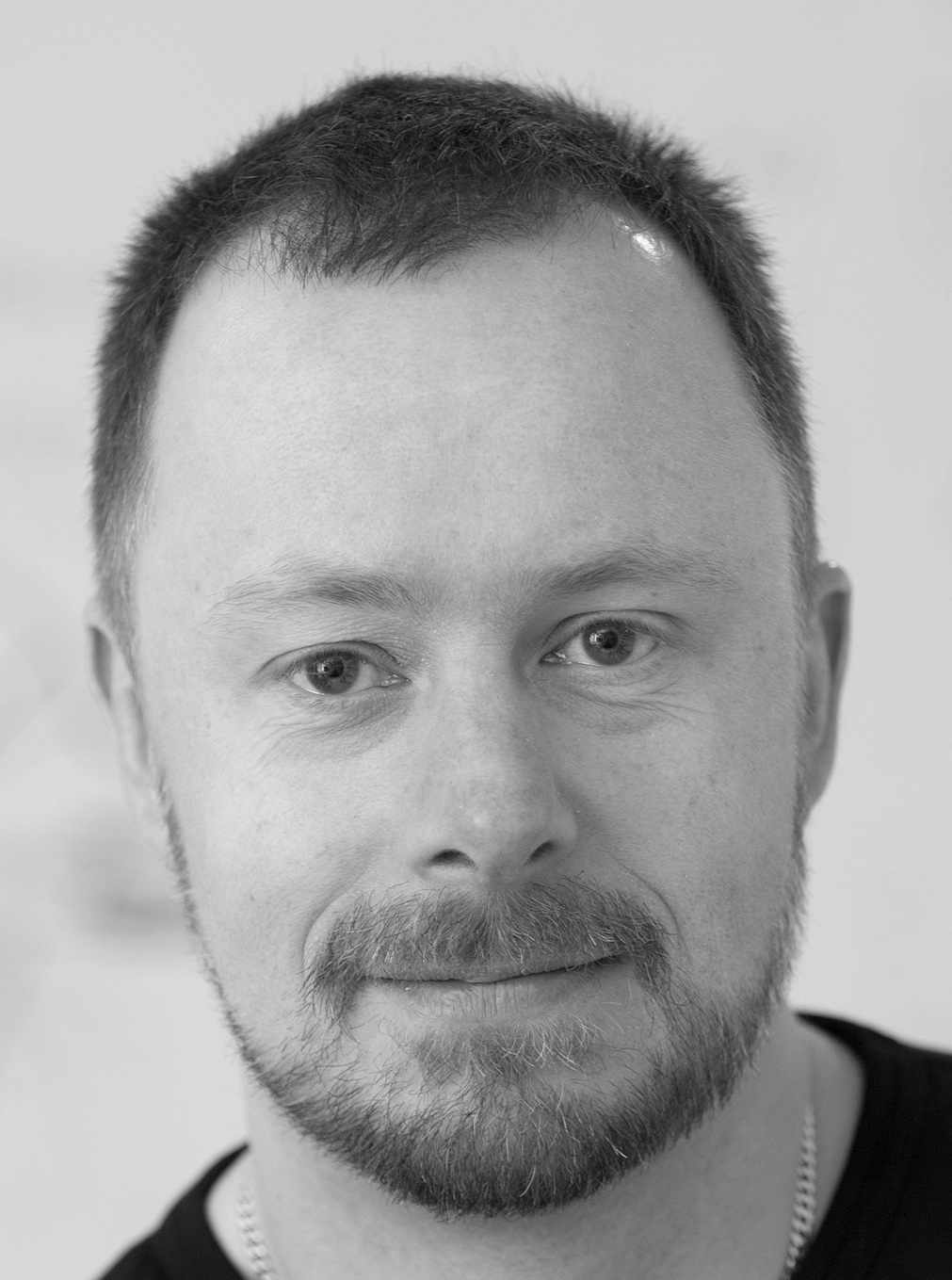}}]{Fredrik Bruzelius}
	received the M.Sc. degree in applied mathematics from Linköping University, Linköping, Sweden, in 1999, and the Ph.D. degree in control theory from the Chalmers University of Technology, Gothenburg, Sweden, in 2004. After a five-year stay in the vehicle industry, he joined the Swedish National Road and Transport Research Institute, Göteborg, Sweden, as a Researcher in vehicle dynamics. Since 2014, he has been an Adjunct Professor with the Chalmers University of Technology. His main research interests include vehicle dynamics estimation and control, tire dynamics, and driving simulators.
\end{IEEEbiography}
\vspace{-1cm}
\begin{IEEEbiography}[{\includegraphics[width=1in,height=1.25in,clip,keepaspectratio]{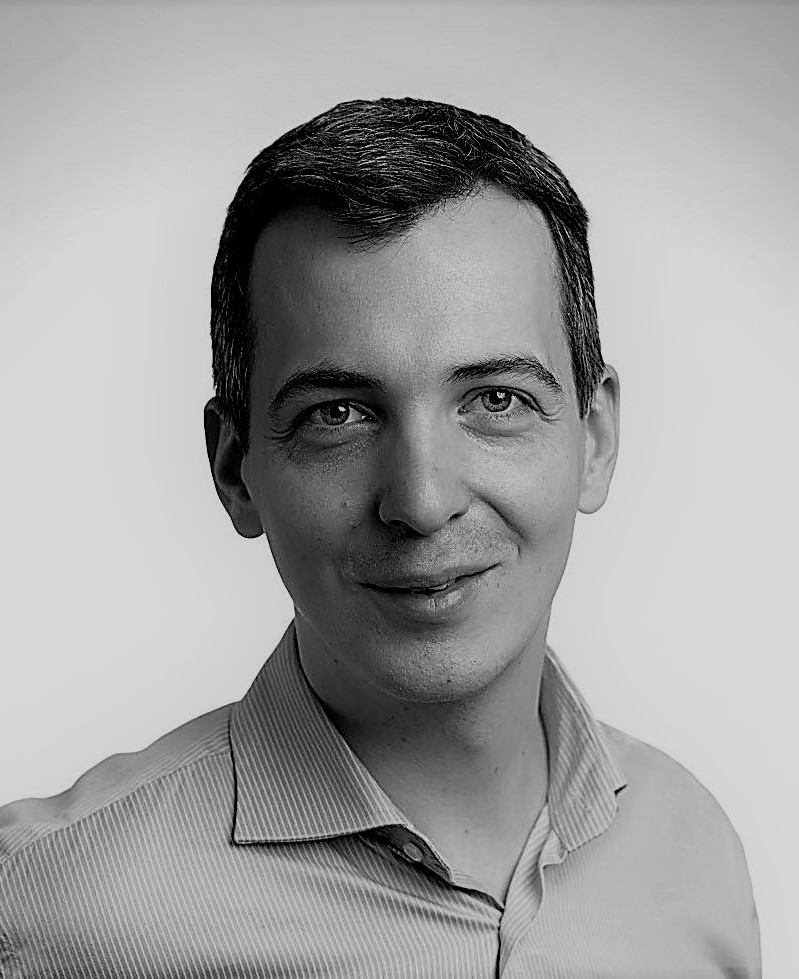}}]{Balázs Kulcsár}
	received the M.Sc. degree in traffic engineering and the Ph.D. degree from Budapest University of Technology and Economics (BUTE), Budapest, Hungary, in 1999 and 2006, respectively. He has been a Researcher/Post-Doctor with the Department of Control for Transportation and Vehicle Systems, BUTE, the Department of Aerospace Engineering and Mechanics, University of Minnesota, USA, and with the Delft Center for Systems and Control, Delft University of Technology, The Netherlands. He is currently a Professor with the Department of Electrical Engineering, Chalmers University of Technology, Sweden. His main research interest focuses on traffic flow modeling and control.
\end{IEEEbiography}

%\vfill

\end{document}